\documentclass{aa}
\usepackage{aas_macros}
\usepackage{natbib}
\usepackage{graphicx}
\usepackage{txfonts}
\usepackage{array}

\def\modif{}

\def\modify{}

\def\wig#1{\mathrel{\hbox{\hbox to 0pt{%
          \lower.6ex\hbox{$\sim$}\hss}\raise.4ex\hbox{$#1$}}}}

\def\aap{{\it Astron. \& Astrophys}}

\begin{document}

\title{Interpreting the yield of transit surveys: \\
Are there groups in the known transiting planets population?}
\titlerunning{Groups within transiting exoplanets?}
\authorrunning{Fressin, Guillot \& Nesta}

\author{Francois Fressin\inst{1}
  \and
  Tristan Guillot\inst{1}
  \and
  Lionel Nesta\inst{2}
  }

\offprints{F. Fressin}

\institute{Observatoire de la C\^ote d'Azur, Laboratoire Cassiop\'ee, CNRS UMR 6202, B.P. 4229, 06304 Nice Cedex 4, France\\
           \email{fressin@obs-nice.fr}
	   \and
Observatoire Fran\c{c}ais des Conjonctures Economiques (OFCE), 250, rue Albert Einstein
06560 Valbonne, France}

\date{A\&A, in press. Received: April 30, 2008, Accepted: December 24, 2008.}

\abstract{Each transiting planet discovered is characterized by 7
measurable quantities, that may or may not be linked together. This
includes those relative to the planet (mass, radius, orbital period, and
equilibrium temperature) and those relative to the star (mass, radius, 
effective temperature, and metallicity). Correlations between planet
mass and period, surface gravity and period, planet radius and star
temperature have been previously observed among the 31 known transiting
giant planets. Two classes of planets have been previously identified based on their
Safronov number.} 
{We use the CoRoTlux transit surveys to compare simulated events to
 the sample of discovered planets and test the statistical
significance of these correlations. Using a model proved to be able to
match the yield of OGLE transit survey, we generate a large sample of
simulated detections, in which we can statistically test the different
trends observed in the small sample of known transiting planets.}  
{We first generate a stellar field with planetary companions based on
  radial velocity discoveries, use a {\modif planetary} evolution model
  assuming a variable fraction of heavy elements {\modif to compute the
    characteristics of transit events}, then apply a detection
  criterion that {\modif includes both statistical and red noise
    sources}. We compare the {\modif yield of our simulated survey}
  with {\modif the ensemble of 31 well-characterized} {\modif giant} transiting
  planets, using different statistical tools, including a multivariate
  logistic analysis to assess whether the simulated distribution
  matches the known transiting planets.}
{Our results satisfactory match the distribution of known transiting
  planets characteristics. {\modif Our multivariate analysis shows that our simulated 
  sample and observations are consistent to 76\%}. The mass vs. period correlation for giant
  planets first observed with radial velocity holds with transiting
  planets. The correlation between surface gravity and period can be
  explained as the combined effect of the mass vs. period lower limit
  and by the decreasing transit probability and detection efficiency
  for longer periods and higher surface gravity. Our model also
  naturally explains other trends, like the correlation between
  planetary radius and stellar effective temperature. Finally, we are
  also able to reproduce the previously observed apparent bimodal
  distribution of planetary Safronov numbers in 10\% of our simulated
  cases, although our model predicts a continuous
  distribution. This shows that the evidence for the existence of two
  groups of planets with different intrinsic properties is not
  statistically significant.}
{}
   

\maketitle
%

\section{Introduction}

The number of giant transiting exoplanets discovered is increasing rapidly
and amounts to 32 at the date of this writing. The ability to measure
the masses and radii of these objects provides us with a unique
possibility to determine their composition and to test planet
formation models. Although uncertainties on stellar and planetary
characteristics do not allow determining the precise composition of
planets individually, a lot is to be learned from a global,
statistical approach.

A particularly intriguing observations made by \citet{Hansen_2007}
from an examination of a set of 18 transiting planets known at that
time is the apparent grouping of objects in two categories based on
their Safronov number.

{\modif The Safronov number $\theta$ is defined as: 
\begin {equation}
\theta = \frac{1}{2} \left[ \frac{V_{esc}}{V_{orb}} \right]^{2} =
\frac{a}{R_p} \frac{M_p}{M_{\star}},
\label{eq:safronov}
\end {equation}
where $V_{esc}$ is the escape velocity from the surface of the planet
and $V_{orb}$ is the orbital velocity of the planet around its host
star, $a$ is the semi-major axis, $M_p$ and $M_{\star}$ are the
respective mass of the planet and its host star, and $R_p$ is the
radius of the planet. It is indicative of the efficiency with which a
planet scatters other bodies, and could play an important role in
understanding processes that affected planet formation.}

If real, this division into two groups
would probably imply the existence of different formation or accretion
mechanisms, or alternatively require revised evolution models. 

Other puzzling observations include the possible trends between planet mass and
orbital period \citep{Mazeh_2005} and between gravity and
orbital period \citep[][ first mentioned by R. Noyes in 2006]{Southworth_2007}.

In a previous article \citep[][--hereafter Paper~I--]{Fressin_2007}, we
presented CoRoTlux, a tool to model statistically a population of
stars and planets and compare it to the ensemble of detected
transiting planets. We showed the results to be in very good agreement
with the 14 planets known at that time. 

In the present article, we examine whether these trends and groups can
be explained in the framework of our model or whether they imply the
existence of more complex physical mechanisms for the formation or
evolution of planets that are not included in present models.  We
first describe our model and an updated global statistical analysis of the
results including 17 new planets discovered thus far
(\S~\ref{sec:update}). We then examine the trends between mass,
gravity and orbital period (\S~\ref{sec:trends}), the {\modif grouping in terms of 
planetary radius and stellar effective temperature (\S~\ref{sec:t_eff}), 
and finally the grouping in terms of Safronov number (\S~\ref{sec:safronov})}.

\section{Method and result update}
\label{sec:update}

\subsection{Principle of the simulations}

As described in more detail in Paper~I, the generation of a population
of transiting planets with CoRoTlux involves the following steps:
\begin{enumerate}
\item We generate a population of stars from the Besan\c{c}on catalog
  \citep{Robin_2003};
\item Stellar companions (doubles, triples) are added using
  frequencies of occurrence and period distributions based on Duquennoy
  \& Mayor \citep{Duquennoy_1991};
\item Planetary companions with random orbital inclinations are
  generated with a frequency of occurrence that depends on the host
  star metallicity with the relation derived by
  \citet{Santos_2004}. The parameters of the planets (period, mass,
  eccentricity) are derived by cloning the known radial-velocity
  (hereafter RV) list of planets (we use J. Schneider's planet
  encyclopaedia: www.exoplanets.eu). We consider only planets above
  $0.3$ times the mass of Jupiter, which yields a list of 229
  objects. This mass cut-off is chosen from radial velocity analysis
  (\citet{Fischer_2005}), as their planetary occurrence law is
  considered unbiased down to this limit. Because of a strong bias of
  transit surveys towards extremely short orbital periods $P$ (less
  than 2 days), we add to the list clones which are drawn from the
  short-orbit planets found from transiting surveys. The probabilities
  are adjusted so that on average $\sim 3$ transit-planet clones
  {\modif with $P\le 2\,$days} are added {\modify to the RV list of
    229 giant planets. This number is obtained by maximum likelihood
    on the basis of the OGLE survey to reproduce both the planet
    populations at very-short periods that are not constrained by
    RV measurements and the ones with longer periods that
    are discovered by both types of surveys} (see Paper~I).
\item We compute planetary radii using a structure and evolution model
  that {\modif is adjusted to fit the radii distribution of known} transiting planets: the
  planetary core mass is assumed to be a function of the stellar
  metallicity, and the evolution is calculated by including an
  extra-heat source term equal to 1\% of the incoming stellar heat
  flux \citep{Guillot_2006, Guillot_2008} \footnote{An electronic
  version of the table of simulated planets used to extrapolate radii
  is available at www.obs-nice.fr/guillot/pegasids/};
\item We determine which transiting planets are detectable, given an
  observational duty cycle and a level of white and red noise
  estimated a posteriori \citep{Pont_2006}. {\modif We also use a cut-off
  in stellar effective temperature $T_{\rm eff, cut}$ above which we
  consider that it will be too difficult for {\modif RV}
  techniques to confirm an event. We choose  $T_{\rm eff, cut}=7200\,$K
  as a fiducial value. {\modify (This value is an estimate of the limit for $T_{\rm eff}$ 
  used by the OGLE follow-up group (F. Pont, pers. communication);
  in practice it has little consequences on the results)}.}
\end{enumerate}

In order to analyze the complete yield of transit discoveries
properly, we should simulate each successful survey (OGLE:
e.g. {\modif see \citet{Udalski_2003}; 
HATnet: e.g. see \citet{Bakos_2006};
TrES: e.g. see \citet{Alonso_2004};
SWASP: e.g. see \citet{Cameron_2006}; 
XO: e.g. see \citet{McCullough_2006})} 
one by one. However, we take advantage of the fact that these
different ground-based surveys have similar observation biases and
similar noise levels (e.g. the red noise level for SWASP
\citep{Smith_2006} is close to the one of OGLE \citep{Pont_2006},
although their instruments and target magnitude range are
different). As a consequence, one can notice that in terms of transit
depth and period distribution of detected transiting planets, these
surveys achieve very similar performances. Therefore, as in Paper~I,
we base our model parameters (stellar fields, duty cycle, red noise
level) on OGLE parameters \citep{Udalski_2003, Bouchy_2004,
  Pont_2005}.

\subsection{The known transiting giant planets}

\begin{table*}[tbp]
\caption{{\modify Characteristics of transiting planets included in this study}}
\label{table:transiting_planets}
\centering
\scriptsize
\begin{tabular}{>{$}l<{$}>{$}r<{$}>{$}r<{$}>{$}r<{$}>{$}r<{$}>{$}r<{$}>{$}r<{$}r} \hline \hline
Name & M_{p} & R_p	& P	& T	& i	& a	& reference \\	
&	[M_{\rm Jup}]	& [R_{\rm Jup}]	& [days] & [JD-2450000] & [^{\circ}] &	[AU] & \\	
\hline \hline	
{\mathbf {HD17156b}}& 3.13 \pm{0.21} &	1.21 \pm{0.12} & 21.21691 \pm{\_71} & 4374.8338 \pm{\_20} & 86.5_{-0.7}^{+1.1} &	0.15 & [Barbieri07]Fischer07/Irwin08 \\
{\mathbf {HD147506b}^{\star \star}} & 8.04 \pm{0.40} &	0.98 \pm{0.04} & 5.63341 \pm{\_13} & 4212.8561 \pm{\_6} & > 86.8 & 0.0685 & [Bakos07]Winn07* \\
{\mathbf {HD149026b}} & 0.36 \pm{0.03} &	0.71 \pm{0.05} & 2.8758882 \pm{\_61} & 4272.7301 \pm{\_13} & 90 \pm{3.1} & 0.0432 & [Sato05]Winn07* \\
{\mathbf {HD189733b}} & 1.15 \pm{0.04} &	1.154 \pm{0.017} &	2.218581 \pm{\_2} &	3931.12048 \pm{\_2} &	85.68 \pm{0.04} &	0.031 & [Bouchy05]Pont07* \\
{\mathbf {HD209458b}} & 0.657 \pm{0.006} &	1.320 \pm{0.025} & 3.52474859 \pm{\_38} &	2826.628521 \pm{\_87} &	86.929 \pm{0.010} &	0.047 & [Charbonneau00]Winn05/Knutson06* \\
{\mathbf {TrES-1}} & 0.76 \pm{0.05} &	1.081 \pm{0.029} & 3.0300737 \pm{\_26} & 3186.80603 \pm{\_28} &	>88.4	& 0.0393 &	[Alonso04]Sozzetti04/Winn07* \\
{\mathbf {TrES-2}} & 1.198 \pm{0.053} &	1.220 ^{+.045}_{-.042} & 2.47063 \pm{\_1} &	3957.6358 \pm{\_10} &	83.90 \pm{0.22} &	0.0367 & [ODonovan06] Sozzetti07* \\
{\mathbf {TrES-3}} & 1.92 \pm{0.23} &	1.295 \pm{0.081} &	1.30619 \pm{\_1} & 4185.9101 \pm{\_3}	& 82.15 \pm{0.21} &	0.0226 &		[ODonovan07]* \\
{\mathbf {TrES-4}} & 0.84 \pm{0.20}	& 1.674 \pm{0.094} &	3.553945 \pm{\_75}	& 4230.9053 \pm{\_5}	& 82.81 \pm{0.33} &	0.0488 &	[Mandushev07]* \\
{\mathbf {XO-1b}} & 0.90 \pm{0.07} &	1.184 ^{+.028}_{-.018} &	3.941534 \pm{\_27} &	3887.74679 \pm{\_15} &	89.36 ^{+.46}_{-.53} & 0.0488 & [McCullough06]Holman06* \\
{\mathbf {XO-2b}} & 0.57 \pm{0.06} &	0.973 ^{+.03}_{-.008} & 2.615838 \pm{\_8}	& 4147.74902 \pm{\_20}	& >88.35 & 0.037 &		[Burke07]* \\
{\mathbf {XO-3b}} & 13.25 \pm{0.64} & 1.1-2.1 & 3.19154 \pm{14} & 4025.3967 \pm{\_38} &	79.32 \pm{1.36} &	0.0476 &	[Johns-Krull08] \\
{\mathbf {HAT-P-1b}}	& 0.53 \pm{0.04} & 1.203 \pm{0.051} &	4.46529 \pm{\_9} &	3997.79258 \pm{\_24} &	86.22 \pm{0.24}	& 0.0551 & [Bakos07]Winn07* \\
{\mathbf {HAT-P-3b}}	& 0.599 \pm{0.026} & 0.890 \pm{0.046} &	2.899703 \pm{54} & 4218.7594 \pm{\_29}	& 87.24 \pm{0.69} &	0.0389  & [Torres07]* \\
{\mathbf {HAT-P-4b}} &	0.68 \pm{0.04} & 1.27 \pm{0.05} & 3.056536 \pm{\_57} & 4245.8154 \pm{\_3} &	89.9 ^{+0.1}_{-2.2} &	0.0446 & [Kovacs07]* \\
{\mathbf {HAT-P-5b}}	& 1.06 \pm{0.11} & 1.26 \pm{0.05}	& 2.788491 \pm{\_25}	& 4241.77663 \pm{\_22} &	86.75 \pm{0.44} &	0.0407 & [Bakos07]* \\
{\mathbf {HAT-P-6b}}	& 1.057 \pm{0.119} & 1.330 \pm{0.061}	& 3.852985 \pm{\_5} & 4035.67575 \pm{\_28} & 85.51 \pm{0.35} &	0.0523 & [Noyes07]* \\
{\mathbf {WASP-1b}} & 0.867 \pm{0.073}	& 1.443 \pm{0.039} & 2.519961 \pm{\_18} & 4013.31269 \pm{\_47} & >86.1 & 0.0382 &	[Cameron06]Shporer06/Charbonneau06* \\
{\mathbf {WASP-2b}} & 0.81-0.95 & 1.038 \pm{0.050}	& 2.152226 \pm{\_4} &	4008.73205 \pm{28} & 84.74 \pm{0.39} & 0.0307 &	[Cameron06]Charbonneau06* \\
{\mathbf {WASP-3b}} & 1.76 \pm{0.11}	& 1.31 _{-.14}^{+.07} &	1.846834 \pm{\_2} & 4143.8503 \pm{\_4} & 84.4 _{-0.8}^{+2.1} &	0.0317 & [Pollacco07] \\
{\mathbf {WASP-4b}} & 1.27 \pm{0.09} &	1.45 _{-.08}^{+.04} &	1.338228 \pm{\_3} & 4365.91475 \pm{\_25} & 87.54 _{-.04}^{+2.3} &	0.023 &	[Wilson08] \\
{\mathbf {WASP-5b}} & 1.58 \pm{0.11} &	1.090 _{-.058}^{+.094} & 1.6284296 \pm{\_42}	& 4375.62466 \pm{\_26} &	> 85.0 & 0.0268 &	[Anderson08] \\
{\mathbf {COROT-Exo-1b}}	& 1.03 \pm{0.12} & 1.49 \pm{0.08}	& 1.5089557 \pm{\_64} & 4159.4532 \pm{\_1} & 85.1 \pm{0.5} & 0.025 & [Barge08] \\
{\mathbf {COROT-Exo-2b}}	& 3.31 \pm{0.16} & 1.465 \pm{0.029}	& 1.7429964 \pm{\_17} & 4237.53562 \pm{\_14} & 87.84 \pm{0.10} & 0.028 & [Alonso08] \\
{\mathbf {OGLE-TR-10b}} & 0.61 \pm{0.13} &	1.122 ^{+0.12}_{-0.07} & 3.101278 \pm{\_4} &	3890.678 \pm{\_1} & 87.2-90 & 0.0416 &	[Konacki05]Pont07/Holman07* \\
{\mathbf {OGLE-TR-56b}} & 1.29 \pm{0.12} &	1.30 \pm{0.05} & 1.211909 \pm{\_1}	& 3936.598 \pm{\_1} & 81.0 \pm{2.2} & 0.0225 & 	[Konacki03]Torres04/Pont07* \\
{\mathbf {OGLE-TR-111b}}	& 0.52 \pm{0.13} & 1.01 \pm{0.04} &	4.0144479 \pm{\_41} & 3799.7516 \pm{\_2} & 88.1 \pm{0.5} & 0.0467 & [Pont04]Santos06/Winn06/Minniti07* \\
{\mathbf {OGLE-TR-113b}}	& 1.32 \pm{0.19} & 1.09 \pm{0.03}	& 1.4324757 \pm{\_13} & 3464.61665\pm{\_10}	& 88.8-90	& 0.0229 &		[Bouchy04]Bouchy04/Gillon06* \\
{\mathbf {OGLE-TR-132}}	& 1.14 \pm{0.12} & 1.18 \pm{0.07}	& 1.689868 \pm{\_3} & 3142.5912 \pm{\_3} & 81.5 \pm{1.6} & 0.0299 &	[Bouchy04]Gillon07* \\
{\mathbf {OGLE-TR-182b}}	& 1.01 \pm{0.15} & 1.13 _{-.08}^{+.24} & 3.97910 \pm{\_1} & 4270.572 \pm{\_2}	& 85.7 \pm{0.3} &	0.051 & [Pont08] \\
{\mathbf {OGLE-TR-211b}}	& 1.03 \pm{0.20} & 1.36 ^{+.18}_{-.09} & 3.67724 \pm{\_3} & 3428.334 \pm{\_3}	& >82.7	& 0.051 &	[Udalski07] \\
    \hline \hline
\multicolumn{8}{l}{\parbox{\textwidth}{{\modify Underscores indicate uncertainties on last printed digits. Bracket = announcement paper. No bracket = reference from which most parameters have been chosen from. *=also in \citet{Torres_2008}.}}}\\
\multicolumn{8}{l}{\parbox{\textwidth}{$\rm M_{\rm Jup}=1.8986112\times 10^{30}\,$g is the
  mass of Jupiter. $\rm R_{Jup}=71,492\,$km is Jupiter's equatorial radius.}}\smallskip\\
\multicolumn{8}{l}{\parbox{\textwidth}{References: \citet{Charbonneau_2000,Konacki_2003,Bouchy_2004,Pont_2004,Torres_2004,Alonso_2004,Sozzetti_2004,Sato_2005,Bouchy_2005,Winn_2005,O_Donovan_2006,Cameron_2006,Knutson_2007,Gillon_2006,Charbonneau_2006,Holman_2006,Shporer_2007,Winn_2007,Winn_2007_b,Bakos_2007,Burke_2007,O_Donovan_2007,Mandushev_2007,Torres_2007,Pont_2008,Gillon_2007,Minniti_2007,Winn_2007_c,Kovacs_2007} \\
The table is derived from Frederic Pont's web site: {\tt http://www.inscience.ch/transits/}.}}\\
\multicolumn{8}{l}{\parbox{\textwidth}{ {\modif $^{\star \star}$ HD147056 is also called HAT-P-2} }}\\
\normalsize
\end{tabular}
\end{table*}

\begin{table*}[tbp]
\caption{{\modify Characteristics of stars hosting the
transiting planets included in this study}}
\label{table:transiting_stars}
\centering
\scriptsize
\begin{tabular}{>{$}l<{$}>{$}r<{$}>{$}r<{$}>{$}r<{$}>{$}r<{$}>{$}r<{$}r} \hline \hline
Name	&	V mag	&	M_{\star}	&	R_{\star}	&	T_{\rm eff}	&	[Fe/H]	&	Reference	\\
& & [M_{\odot}] & [R_{\odot}] & [K] & & \\
\hline \hline
{\mathbf {HD17156}}	&	8.2	&	1.2 \pm{0.1}	&	1.47 \pm{0.08}	&	6079 \pm{56}	&	0.24 \pm{0.03}	&	Fischer07/Irwin08	\\
{\mathbf {HD147506}} 	&	8.7	&	1.32 \pm{0.08}	&	1.48 \pm{0.05}	&	6290 \pm{110}	&	0.12 \pm{0.08}	&	Bakos07/Winn07*	\\
{\mathbf {HD149026}} 	&	8.2	&	1.3 \pm{0.06}	&	1.45 \pm{0.1}	&	6147 \pm{50}	&	0.36 \pm{0.05}	&	Sato05/Winn07*	\\
{\mathbf {HD189733}} 	&	7.7	&	0.82 \pm{0.03}	&	0.755 \pm{0.011}	&	5050 \pm{50}	&	-0.03 \pm{0.04}	&	Bouchy05/Pont07*	\\
{\mathbf {HD209458}} 	&	7.7	&	1.101 \pm{0.064}	&	1.125 \pm{0.022}	&	6117 \pm{26}	&	0.02 \pm{0.03}	&	Sozzetti04/Knutson06*	\\
{\mathbf {TrES-1}} 	&	11.8	&	0.89 \pm{0.035}	&	0.811 \pm{0.020}	&	5250 \pm{75}	&	-0.02 \pm{0.06}	&	Sozzetti04,06/Winn07*	\\
{\mathbf {TrES-2}} 	&	11.4	&	0.98 \pm{0.062}	&	1.000 _{-.033}^{+.036}	&	5850 \pm{50}	&	-0.15 \pm{0.10}	&	Sozzetti07*	\\
{\mathbf {TrES-3}} 	&	12.4	&	0.90 \pm{0.15}	&	0.802 \pm{0.046}	&	5720 \pm{150}	&		&	O Donovan07*	\\
{\mathbf {TrES-4}} 	&	11.6	&	1.22 \pm{0.17}	&	1.738 \pm{0.092}	&	6100 \pm{150}	&		&	Mandushev07*	\\
{\mathbf {XO-1}} 	&	11.5	&	1.0 \pm{0.03}	&	0.928 \pm{0.015}	&	5750 \pm{13}	&	0.015 \pm{0.03}	&	MCCullough06/Holman06*	\\
{\mathbf {XO-2}} 	&	11.2	&	0.98 \pm{0.02}	&	0.964 ^{+.02}_{-.009}	&	5340 \pm{32}	&	0.45 \pm{0.02}	&	Burke07*	\\
{\mathbf {XO-3}} 	&	9.8	&	1.41 \pm{0.08}	&	{\modif 1.377 \pm{0.083}}	&	6429 \pm{50}	&	-0.18 \pm{0.03}	&	Johns-Krull07	\\
{\mathbf {HAT-P-1}}	&	10.4	&	1.12 \pm{0.09}	&	1.115 \pm{0.043}	&	5975 \pm{45}	&	0.13 \pm{0.02}	&	Bakos07/Winn07*	\\
{\mathbf {HAT-P-3}}	&	11.9	&	0.936 ^{+.036}_{-.062}	&	0.824 ^{+.043}_{-.035}	&	5185 \pm{46}	&	0.27 \pm{0.04}	&	Torres07*	\\
{\mathbf {HAT-P-4}} 	&	11.2	&	1.26 ^{+.06}_{-.14}	&	1.59 \pm{0.07}	&	5860 \pm{80}	&	0.24 \pm{0.08}	&	Kovacs07*	\\
{\mathbf {HAT-P-5}}	&	12.0	&	1.160 \pm{0.062}	&	1.167 \pm{0.049}	&	5960 \pm{100}	&	0.24 \pm{0.15}	&	Bakos07*	\\
{\mathbf {HAT-P-6}}	&	10.4	&	1.29 \pm{0.06}	&	1.46 \pm{0.06}	&	6570 \pm{80}	&	-0.13 \pm{0.08}	&	Noyes07*	\\
{\mathbf {WASP-1}} 	&	11.8	&	1.15 ^{+.24} _{-.09}	&	1.453 \pm{0.032}	&	6110 \pm{45}	&	0.23 \pm{0.08}	&	Cameron06 Charbonneau06/Stempels07*	\\
{\mathbf {WASP-2}} 	&	12.0	&	0.79 ^{+.15}_{ -.04}	&	0.813 \pm{0.032}	&	5200 \pm{200}	&		&	Cameron06/Charbonneau06*	\\
{\mathbf {WASP-3}} 	&	10.5	&	1.24 _{-.11}^{+.06}	&	1.31 _{-.12}^{+.05}	&	6400 \pm{100}	&	0.0 \pm{0.2}	&	Pollacco07	\\
{\mathbf {WASP-4}} 	&	12.5	&	0.90 \pm{0.08}	&	0.95 _{-.03}^{+.05}	&	5500 \pm{150}	&	0.0 \pm{0.2}	&	Wilson08	\\
{\mathbf {WASP-5}} 	&	12.3	&	0.97 \pm{0.09}	&	1.026 _{-.044}^{+.073}	&	5700 \pm{150}	&	0.0 \pm{0.2}	&	Anderson08	\\
{\mathbf {COROT-Exo-1}}	&	13.6	&	0.95 \pm{0.15}	&	1.11 \pm{0.05}	&	5950 \pm{150}	&	-0.3 \pm{0.25}	&	Barge08	\\
{\mathbf {COROT-Exo-2}}	&	12.57	&	0.97 \pm{0.06}	&	0.902 \pm{0.018}	&	5625 \pm{120}	&	\sim 0	&	Alonso08/{\modify Bouchy08}	\\
{\mathbf {OGLE-TR-10}} 	&	15.8	&	1.10 \pm{0.05}	&	1.14 ^{+0.11}_{-0.6}	&	6075 \pm{86}	&	0.28 \pm{0.10}	&	Santos06/Pont07*	\\
{\mathbf {OGLE-TR-56}} 	&	16.6	&	1.17 \pm{0.04}	&	1.32 \pm{0.06}	&	6119 \pm{62}	&	0.19 \pm{0.07}	&	Santos06/Pont07*	\\
{\mathbf {OGLE-TR-111}}	&	17.0	&	0.81 \pm{0.02}	&	0.83 \pm{0.03}	&	5044 \pm{83}	&	0.19 \pm{0.07}	&	Santos06/Winn06*	\\
{\mathbf {OGLE-TR-113}}	&	16.1	&	0.78 \pm{0.02}	&	0.77 \pm{0.02}	&	4804 \pm{106}	&	0.15 \pm{0.10}	&	Santos06/Gillon06*	\\
{\mathbf {OGLE-TR-132}}	&	16.9	&	1.26 \pm{0.03}	&	1.34 \pm{0.08}	&	6210 \pm{59}	&	0.37 \pm{0.07}	&	Gillon07*	\\
{\mathbf {OGLE-TR-182}}	&	17	&	1.14 \pm{0.05}	&	1.14 _{-.06}^{+.23}	&	5924 \pm{64}	&	0.37 \pm{0.08}	&	Pont08	\\
{\mathbf {OGLE-TR-211}}	&	15.5	&	1.33 \pm{0.05}	&	1.64 ^{+.21}_{-.05}	&	6325 \pm{91}	&	0.11 \pm{0.10}	&	Udalski07	\\
\hline \hline
\multicolumn{7}{l}{\parbox{\textwidth}{Underscores indicate uncertainties on last printed digits. *=also in \citet{Torres_2008}}}\\
\multicolumn{7}{l}{\parbox{\textwidth}{References: \citet{Charbonneau_2000,Konacki_2003,Bouchy_2004,Pont_2004,Torres_2004,Alonso_2004,Sozzetti_2004,Sato_2005,Bouchy_2005,Winn_2005,O_Donovan_2006,Cameron_2006,Knutson_2007,Gillon_2006,Charbonneau_2006,Holman_2006,Shporer_2007,Winn_2007,Winn_2007_b,Bakos_2007,Burke_2007,O_Donovan_2007,Mandushev_2007,Torres_2007,Pont_2008,Gillon_2007,Minniti_2007,Winn_2007_c,Kovacs_2007,Bouchy_2008}; The discovery Papers are in brackets. The table is taken from F. Pont's site: {\tt http://www.inscience.ch/transits/}.}}
\normalsize
\end{tabular}
\end{table*}

Our results will be systematically compared to the sample of 31
transiting giant planets that are known at the date of this
writing. These include in particular: 
\begin{itemize}
	\item 22 planets for which the refined parameters based on the
uniform analysis of transit light curves and the observable properties
of the host stars have been generically updated by
\citet{Torres_2008}. {\modif We exclude the {\modify sub-giant} Hot Neptune GJ-436 b that does
not fit our mass criterion and is undetectable by current 
ground-based generic surveys.}
	\item 9 planets recently discovered {\modif and not included in \citet{Torres_2008}}. 
	The characteristics of
these planets have not been refined and are to be considered with more
caution. Among these planets, we added the first two discoveries of
the CoRoT satellite. Although CoRoT has significantly higher
photometric precision and {\modify is better suited for finding longer period planets than}
ground based surveys, we included both CoRoT-Exo-1b
\citep{Barge_2008} and CoRoT-Exo-2b \citep{Alonso_2008} in our
analysis, as they are the two deepest planets candidates of the
initial run of the satellite and have similar periods and transit
depths to planets discovered from ground-based surveys.
\end{itemize}
The characteristics of the transiting planets are shown in
Table~\ref{table:transiting_planets} for transiting planets
characteristics and Table~\ref{table:transiting_stars} for their host
stars. These tables are used for testing our model. 
{\modify Where the stellar metallicity is unknown, we arbitrarily used solar metallicity
(see below and the appendix for a discussion)}.

\subsection{{\modif A new metallicity distribution for stars hosting planets}}

In Paper~I, we had concluded that the metallicity distribution of
stars with Pegasids ({\modif planets with masses between $0.3$ and $15 M_{\rm Jup}$ and periods
  $P<10$\,days}) was significantly different from those of stars with planets having longer
orbital periods. This was based on three facts:
\begin{itemize}
\item The list of radial-velocity planets known showed a lack of giant
  planets with short orbital periods around metal-poor stars. Among 25
  Pegasids, none were orbiting stars with $\rm
  {\rm [Fe/H]}<-0.07$, contrary to planets on longer orbits found also
  around metal-poor stars.
\item The list of transiting planets also showed a lack of planets
  around metal-poor stars, with stellar metallicities ranging from
  $-0.03$ to $0.37$ ($[-0.08,0.44]$ with error bars).
\item The population of transiting planets generated with CoRoTlux
  was found to systematically underpredict stellar
  metallicities compared to the sample of observed transiting
  planet. The period vs. metallicity diagram thus formed was found to be
  $2.9\sigma$ away from the maximum likelihood of {\modif simulated planets position
  in the diagram (see Paper I)}.\footnote{Paper~I shows how we estimate the deviation of real 
  planets from maximimum likelihood of the model: in each 2-parameter space, we bin our data on a 20x20 grid as a
  compromise between resolution of the models and characteristic variations of the parameters.
  The probability of an event in each bin is considered equal to the normalized number of
  draws in that bin in our large model sample. The likelihood of a 31-planets draw is the sum of the logarithms 
  of the individual probablities of its events. We estimate the standard deviation of 1000 random 31-events draws among 
  the model detections sample, and calculate the deviation to maximum likelihood of the known planets as a function of this standard deviation.}

  On the other hand, a similar calculation done by splitting the {\modify RV} list
  in a low-metallicity part {\modif (${\rm [Fe/H]}<-0.07$)} and a high-metallicity
  part {\modif (with two different period distribution for simulated planets as a function of their host star metallicity)} would end in a period vs. metallicity diagram in good agreement
  with the observations ($0.4\sigma$ from the maximum likelihood).  
\end{itemize}

On the basis of an additional 51 RV giant planets and {\modif 17}
transiting planets discovered {\modif since Paper~I}, we must now
reexamine this conclusion. Indeed, the average metallicity of stars
harboring transiting planets has evolved. The OGLE survey was
characterized by a surprisingly high value (${\rm [Fe/H]}=0.24$). The
planets discovered since have significantly lower metallicities (an
average of ${\rm [Fe/H]}=0.07$). Finally, {\modify TrES-2, TrES-3}, XO-3, HAT-P-6 and
{\modify CoRoT-Exo-1} all appear to have metallicities lower than $-0.07$.

\begin{figure}
\centerline{\resizebox{7cm}{!}{\includegraphics{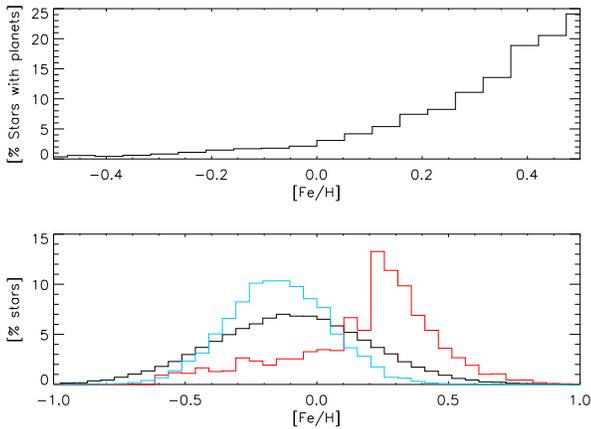}}}
\vspace*{1cm}
\caption{Distribution of stars as a function of their metallicity
  {\rm [Fe/H]}. {\it Upper panel\/}: Fraction of stars with planets as a
  function of their metallicity, as obtained from radial velocity surveys
  \citep{Santos_2004, Fischer_2005}. {\it Bottom panel\/}: Normalized
  distribution of stellar metallicities assumed in Paper~I (blue) and
  in this work (black). The resulting {\rm [Fe/H]} distribution of
  planet-hosting stars is also shown in red.}
\label{fig:metallicities}
\end{figure}

In Paper~I, the metallicity distribution of simulated stars was based on that
extracted from the photometric observation of solar neighborhood of
the Geneva-Copenhagen survey \citep{Nordstrom_2004}. This metallicity
distribution is in fact centred one dex lower ($-0.14$ instead of
$-0.04$) than the one observed using spectrometry by RV surveys
\citep{Fischer_2005, Santos_2004}. Since {\modify the latter two} works are used to
derive the frequency of stars bearing planets, we now choose to also
use these for the metallicity distribution of stars in our
fields. More specifically, our metallicity distribution law and the
planet occurrence rate are obtained by combining the
\citet{Santos_2004} and the \citet{Fischer_2005} surveys.  
Figure~\ref{fig:metallicities} shows the metallicity distribution and
planet occurrence that result directly from these hypotheses. 

As a consequence, we find that with this improved distribution of
stellar metallicities with the new sample of observed planets
alleviates the need for advocating a distinction in metallicities
between stars harboring short-period giant planets and stars that
harbor planets on longer periods. Quantitatively, our new metallicity
vs. period diagram is at $1.09\sigma$ of the maximum likelihood. We
therefore conclude that, contrary to Paper~I, there is no
statistically significant bias between the planet periodicity and the
stellar metallicity in the observed exoplanet sample.

\subsection{Statistical evaluation of the performances of the model}
\label{sec:stat}

{\modif As shown in detail in the appendix (see online version), the
  model is evaluated using univariate, two-dimensional and
  multivariate statistical tests. Specifically, we show that the
  parameters for the simulated and observed planets have globally the
  same mean and standard deviation and that both Student-t tests and
  Kolmogorov-Smirnov tests indicate that the two populations are
  statistically indistinguishable. However, while these univariate
  tests provide preliminary tests of the quality of the data,
  they are not sufficient because of the multiple correlations between parameters
  of the problem. 
}

\begin{table*}
\caption{Pearson correlations between planetary and stellar characteristics. 
Significant correlations ($\ge 0.5$) are boldfaced.}\label{table:all_correlations}\centering\begin{tabular}{ccccccccccc}\hline\hline& {\modif $Y^{\star}$} & $\theta$ & ${\rm [Fe/H]}$ & $T_{\rm eff}$ & $R_{\star}$ & $M_{\star}$ & $T_{eq}$ & {\modify $P$} & $R_p$ & $M_p$ \\
\hline $\theta$ & $-0.0046$ & &&&&&&&&\\
${\rm [Fe/H]}$ & $0.0048$ & $0.0560$ &&&&&&&\\
$T_{\rm eff}$ & $-0.0013$ & $-0.1300$ & $-0.0227$ &&&&&&\\
$R_{\star}$ & $0.0006$ & $0.1240$ & $-0.0283$ & {$\mathbf {0.8103}$} &&&&&\\
$M_{\star}$ & $0.0003$ & $-0.1301$ & $-0.0237$ & {$\mathbf {0.9359}$} & {$\mathbf {0.8761}$} &&&&\\
$T_{eq} $ & $0.0006$ & $-0.1411$ & $-0.0013$ & {$\mathbf {0.7338}$} & {$\mathbf {0.7605}$} & {$\mathbf {0.7380}$} &&&\\
{\modify $P$} & $-0.0029$ & $0.4184$ & $0.0068$ & $-0.0304$ & $-0.0280$ & $-0.0303$ & $-0.3990$ &&\\
$R_p$ & $0.0015$ & $-0.3038$ & {$\mathbf {-0.5129}$} & {$\mathbf {0.4833}$} & {$\mathbf {0.5203}$} & {$\mathbf {0.5030}$} & {$\mathbf {0.5191}$} & $-0.1931$ &\\
$M_p$ & $-0.0046$ & {$\mathbf {0.6560}$} & $0.0676$ & $-0.0317$ & $-0.0318$ & $-0.0324$ & $-0.0476$ & $-0.3250$ & $-0.1457$\\
\hline\hline
\multicolumn{11}{l}{{\modif $\star$: variable $Y$ has value 1 if the planet is observed, 0 if it is simulated}.}
\end{tabular}
\end{table*}

{\modif Table~\ref{table:all_correlations} presents the Pearson
  correlation coefficients between each variable. 
 It shows that
  the problem indeed possesses multiple, complex correlations. 
  In this table, the variable $Y$ characterizes the `reality' of the planet
  considered (it is equal to $1$ if the planet of the list is an
  observed one, and to $0$ if it is a simulated planet). We see
  that $Y$ is very weakly correlated with parameters of the
  problem. This indicates that the model is well-behaved, but does not
  constitute a complete validity test in itself.

Table~\ref{table:logitres} presents the results of a multivariate test using a so-called logistic regression (see the appendix for more details). This method allows to include simultaneously all planet characteristics as predictors of the probability of being a known transiting planet (hereafter named `real' planets as opposed to simulated ones), thereby controlling for the correlations between variables at once. Based on maximum likelihood estimation method, it provides information on whether a given characteritics is positively (resp. negatively) and significantly (resp. non significantly) related with the fact of being a real planet. Moreover, it computes the probability ${\cal{P}}_{\chi^2}$ as a general assessment of the quality of the fit. In our case, a large ${\cal{P}}_{\chi^2}$ implies no significant difference between the simulated and real planets. Globally, the general fit of the model shows that simulated planets are not significantly distinct from real planets (${\cal{P}}_{\chi^2}= 0.765$). This can be compared to a model in which model radii are artificially increased by 10\%, for which ${\cal P}_{\chi^2}\sim 10^{-4}$ (see appendix)}

\begin{table}[htbp]
\caption{Logistic maximum likelihood estimates: {\modify $\hat\beta$ is indicative of a correlation with $Y$; ``t-stat'' is the
  the distance in standard deviations from no correlation, and ${\cal{P}}$ represents the probability that the
  model and observations are not significantly different.}}
 \label{table:logitres}
 \centering
 \begin{tabular}{cccc}
 \hline
 \hline Variable & $\hat{\beta}$& t-stat.& ${\cal{P}}$ \\
 \hline
$M_{\star}$  &     0.467 &     0.63 &     0.528  \\
 ${\rm [Fe/H]}$  &      0.415 &     1.39 &     0.164  \\
$T_{\rm eff}$  &     -0.517 &     -0.81 &     0.417  \\
$R_{\star}$  &     0.059 &     0.12 &     0.901  \\
      $P$  &     -0.235 &     -0.32 &     0.746  \\
    $M_p$  &      0.329 &     0.35 &     0.726  \\
    $R_p$  &      0.305 &     0.90 &     0.370  \\
 $T_{eq}$  &     -0.296 &     -0.46 &     0.648  \\
 $\theta$  &     -0.904 &     -0.58 &     0.563  \\
 \hline\hline
 \multicolumn{4}{l}{Maximum likelihood estimations}\\
\multicolumn{4}{l}{Probability ${\cal{P}}_{\chi^2}$= 0.756}\\
\end{tabular}
\end{table}

{\modif Table~\ref{table:logitres} also presents for each seven independent variables
  of the problem plus the planet equilibrium temperature $T_{eq}$ and Safronov number $\theta$ 
  how a given variable is correlated with the fact that a planet is ``real'' (as
  opposed to being one of the simulated planets in the list). The
  different statistical parameters presented in this table are defined
  in the appendix. We only provide here a short
  description: $\hat\beta$ is indicative of a correlation between a
  given variable and the $Y$ (reality) variable. ``t-stat'' represent
  the distance from the mean in terms of standard deviations
  (student-t test). ${\cal{P}}$ represents the probability that the
  correlation is significant. The two last parameters are evaluated
  using bootstrap. 

{\modify The fact that the parameters $\hat\beta$ in
  table~\ref{table:logitres} are non-zero indicate that there is a
  correlation between each parameter and the variable $Y$. However,
  the t-student test indicates that in every case but one (for
  [Fe/H]), the values obtained for $\hat\beta$ are consistent with $0$
  to within one standard-deviation: the agreement between model and
  observations is good. This is further shown by} the high ${\cal{P}}$
values (indicative of consistency between model and observations): The
lowest ${\cal{P}}$ value is associated with the stellar
  metallicty ${\rm [Fe/H]}$, but it is high enough not to show a
  statistically significant difference between our modeled sample and
  real observations. However, this characteristic is the one with the
  largest error bars, and the only one to have missing data (for
  TrES-3, TrES-4, WASP-2 and CoRoT-Exo-2). We included ${\rm [Fe/H]}$,
  as it is an important feature of our model, in our multivariate
  analysis, but the comparison with real planets for this
  characteristic is to be considered carefully. The quality of the
  agreement between observed planets characteristics and our model
  improves to $88.4 \%$ if we remove ${\rm [Fe/H]}$ from our logistic
  maximum likelihood estimates (see the appendix for details and
  further tests).}

  


  

\subsection{Updated mass-radius diagram}

Throughout the article, we will use density maps of the simulated
detections and compare them to the observations. These density maps
use a resolution disk template to get smooth plots. The size of the
resolution template is a function of the number of events present in
the diagram. The color levels follow a linear density rule for most
diagrams we show. In the case of specific diagrams showing rare long
period discoveries (more than 5 days) and large surface gravity or
Safronov number, we choose to use a logarithmic color range for
density maps to emphasize these rare events. A probability map is
established using the {\modif model detections sample (50,000 detections 
obtained by simulating multiple times {\modify the number of observations from the} OGLE survey). Again, we stress that we limited our model to planets below $0.3 M_{\rm Jup}$, both because the question of the composition becomes more  
important and complex for small planets, and because RV detection biases are also more significant. 
their distribution is only partially known from RV surveys}.

\begin{figure}
\centerline{\resizebox{9cm}{!}{\includegraphics{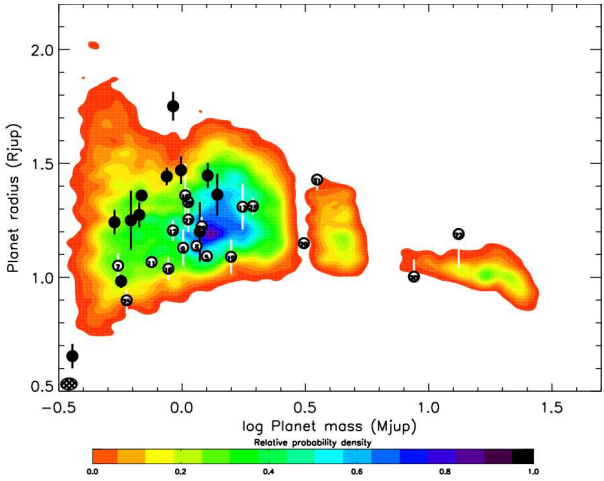}}}
\caption{Mass - radius relation for the transiting Pegasids discovered
  to date (filled circles for planets with low Safronov number $\theta
  < 0.05$, open circles for planets with higher $\theta$
  values). {\modif The joint probability density map obtained from our
    simulation is shown as grey contours (or color contours in the
    electronic version of the article). The resolution disk size used
    for the contour plot appears in the bottom left part of the
    picture. At a given (x,y) location the normalized joint
    probability density is defined as the number of detected planets
    in the resolution disk centered on (x,y) divided by the maximum
    number of detected planets in a resolution disk anywhere on the
    figure.}
}
\label{fig:mass_radius_map}
\end{figure}

Figure~\ref{fig:mass_radius_map} shows the mass-radius diagram density
map simulated with CoRoTlux and compared to the known planets. Gaps in
the diagram at $\sim 3 M_{\rm Jup}$ and $\sim 6-7 M_{\rm Jup}$ are due to the
small sample of close-in RV planets in these ranges and the fact that
our mass distribution is obtained by cloning these observed planet
rather than relying on a smooth distribution (see Paper~I for a
discussion). {\modif These gaps} should disappear with more discoveries
of close-in planets by RV.  Otherwise, the model distribution and the
known planets are in fairly good agreement, as indicated by the
$1.7\sim 1.8\sigma$ distance to the maximum likelihood for this
diagram (Table~\ref{table:likelihood}). However, the agreement is not
{\modif as good as one would expect probably because of two planets
  that possess especially large radii CoRoT-Exo-2b and TrES-4b. The
  existence of these planets is a problem for evolution models in
  general that goes beyond the present statistical tests that we
  propose in this article.}

\section{Trends between mass, surface gravity and orbital period}
\label{sec:trends}

\subsection{A correlation between mass and orbital period of Pegasids}
\label{sec:mass period}

\begin{figure}[htbp]
\centering
\includegraphics[angle=0,width=1.6\columnwidth]{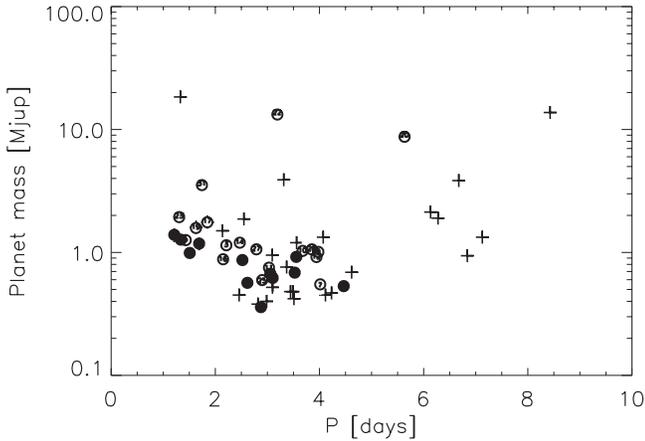}
\caption{Mass-period distribution of known short-period
  exoplanets. Crosses correspond to {\modif non-transiting planets
    discovered by radial-velocity surveys}. Open and filled circles
  correspond to transiting planets (with Safronov numbers below and
  over $0.05$, respectively)}
\label{fig:period_transits_RV}
\end{figure}

Figure~\ref{fig:period_transits_RV} compares the known radial-velocity
planets to the ones detected in transit. {\modif The figure highlights
  the fact that transit surveys are clearly biased towards detecting
  short-period planets. However, as shown in Paper~I and furthermore
  reinforced in the present study,} the
two populations are perfectly compatible provided a limited proportion
of very small planets ($P < 2 $\,days) are added. 

\citet{Mazeh_2005} had pointed out the possibility of an intriguing
correlation between the masses and periods of the six first known
transiting exoplanets. Figure~\ref{fig:period_transits_RV} shows that
the trend is confirmed with the present sample of planets. This
correlation may be due to a migration rate that is inherently
dependant upon planetary mass or to other formation mechanisms. It is
not the purpose of the present article to analyze this
correlation. However, because we use clones of the radial-velocity
planets in our model, it is important to stress that this absence of
small-mass planets with very short orbital distances can subtend some
of the results that will be discussed hereafter.  

\subsection{A correlation between surface gravity and orbital period of Pegasids ?}

\begin{figure}
\centerline{\resizebox{9cm}{!}{\includegraphics{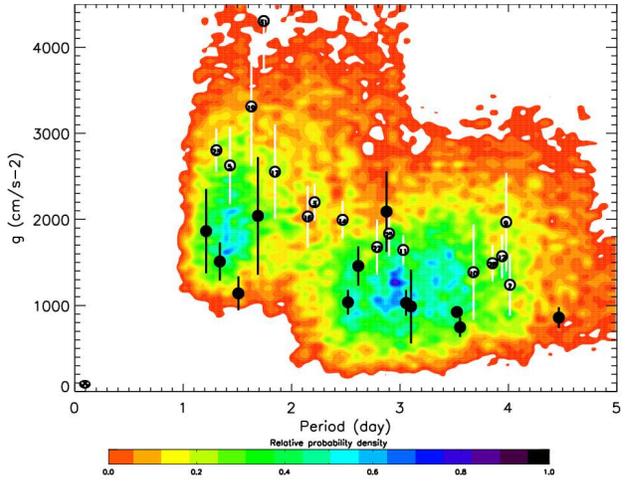}}}
\caption{{\modif Planetary surface gravity versus orbital period of
    transiting giant planets discovered to date (circles) compared to
    a simulated joint probability density map (contours). {\modify Symbols and density plot are the same as in fig.~\ref{fig:mass_radius_map}}.}}
\label{fig:surface_gravity}
\end{figure}

\begin{figure}
\centerline{\resizebox{9cm}{!}{\includegraphics{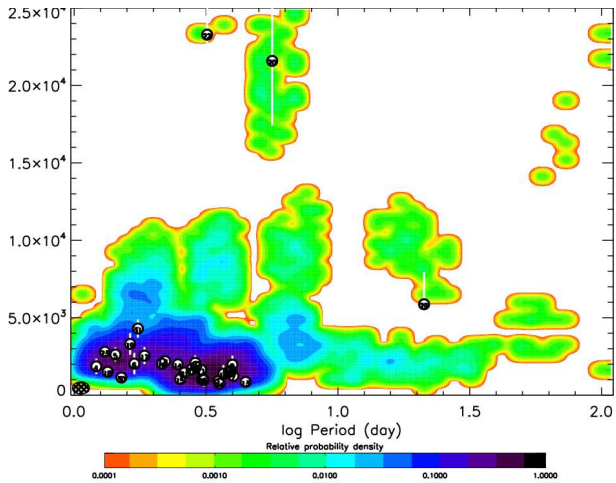}}}
\caption{Same figure as Figure~\ref{fig:surface_gravity} with extended surface gravity and period ranges. Note that the scale of the color levels is logarithmic, in order to emphasize the presence of outliers.}
\label{fig:surface_gravity_extended}
\end{figure}

The existence of a possible anti-correlation between planetary surface
gravity $g=G M_{\rm p} / R_{\rm p}^2$ and the orbital period of the
nine first transiting planets has been pointed out for some time
\citep{Southworth_2007}. This correlation still holds
(fig.~\ref{fig:surface_gravity}) for the Pegasids with periods below 5
days and with jovian masses discovered to date. \modif{At the same
  time, it is important to stress that massive objects (XO-3b,
  HAT-P-2b and HD17156b) are clear outliers (see
  fig.~\ref{fig:surface_gravity_extended})}: {\modify Their much larger
  surface gravity probably implies that they are in a different regime}.

Our model agrees well with the observations ({\modif in this $P-g$
  diagram} real planets are at $0.51$ $\sigma$ from maximum likelihood
of the simulated results).  We can explain the apparent correlation in
Figure~\ref{fig:surface_gravity} {\modif as stemming from the
  existence of two zones with few detectable transiting giant
  planets}:
\begin{enumerate}
	\item The bottom left part of the diagram where planets are
          rare, because of a lack of light planets {\modif(with low
            surface gravity)} with short periods, as discussed in
          section~\ref{sec:mass period};
	\item The upper right part of the diagram (high surface
          gravity, low planetary radius) where transiting planets are
          less likely to transit and more difficult to detect.
\end{enumerate}

Figure~\ref{fig:surface_gravity_extended} shows {\modif the same
  probability density map as in fig.~\ref{fig:surface_gravity} but at
  a larger scale in period and gravity. The three outliers to the
  ``correlation'' appear. These are the
  large mass planets XO-3b, HAT-P-2b and HD17156b. Given the method
  chosen to draw the planet population with CoRoTlux, the probability
  density function that we derive is small, but non-zero around these
  objects, and also elsewhere in the diagram due to the presence of
  non-transiting giant planets with appropriate characteristics. Seen
  at this larger scale, it is clear that the planetary-gravity
  vs. period relation is much more complex than a simple linear
  relation.}

{\modif Globally, figures~\ref{fig:surface_gravity} and
  \ref{fig:surface_gravity_extended} indicate that the relation
  between planet surface gravity and orbital period is {\it not} a
  consequence of a link between the planet {\it composition} and its
  orbital period. Rather, we see it as a consequence of the
  correlation between planetary mass and orbital period for short
  period giant planets, which is, as discussed in the previous
  section, probably linked to mass-dependent migration mechanisms.}

\section{A correlation between stellar effective temperature and planet radius ?}
\label{sec:t_eff}
\label{sec:radius_t_eff}

{\modif The range of radius of Pegasids is surprisingly large,
  especially when one considers the difference in compositions (masses
  of heavy elements varying from almost 0 to $\sim 100\,\rm M_\oplus$)
  that are required to explain known transiting planets within the
  same model \citep{Guillot_2006, Guillot_2008}. Our underlying planet
  composition/evolution model is based on the assumption of a
  correlation of the stellar metallicity with the heavy element
  content in the planet. We checked that no other variable is
  responsible for a correlation that would affect this conclusion.

  We present the results obtained in the $T_{\rm eff}-R_{\rm p}$
  diagram as they are the most interesting: the two variables indeed
  are positively correlated. Furthermore, given that errors in the stellar
  parameters are the main sources of uncertainty in the planetary
  radii determinations, one could suppose that a systematic error in
  the stellar radius measurement as a function of its effective temperature 
  could be the cause of the
  variation in the estimated planetary radii. If true, this may
  alleviate the need for extreme variations in composition. It would
  cast doubts on the stellar metallicity vs. planetary heavy elements
  content correlation. 

\begin{table}
\caption{Mean planet radius for cool versus {\modif hot} stars}
\label{table:mean_radius}
\centering
\begin{tabular}{ccc}
\hline\hline
 & Cool stars & Hot stars \\
 & $T_{\rm eff}< 5400$\,K & $T_{\rm eff} \ge 5400$\,K\\
\hline
``Real'' planets & $1.072 R_{\rm Jup}$ & $1.267 R_{\rm Jup}$ \\
\hline
{\modify All simulated planets} & $1.058 R_{\rm Jup}$ & $1.202 R_{\rm Jup}$ \\
{\modify Detectable simulated planets} & $1.074 R_{\rm Jup}$ & $1.251 R_{\rm Jup}$\\
\hline\hline
\end{tabular}
\end{table} 

  As shown in Table~\ref{table:mean_radius}, the mean radius of
  planets orbiting cool stars ($T_{\rm eff} < 5400 K$) is $1.072\rm R_{\rm
    Jup}$ and it is $1.267\rm R_{\rm Jup}$ for planets orbiting hot
  stars ($T_{\rm eff} \ge 5400 K$). Slightly smaller values are obtained
  in our simulation when considering {\it all} transiting
  planets. However, the values obtained when considering only the {\it
    detectable} transiting planets are in extremely good agreement
  with the observations. 

\begin{figure}
\centerline{\resizebox{9cm}{!}{\includegraphics{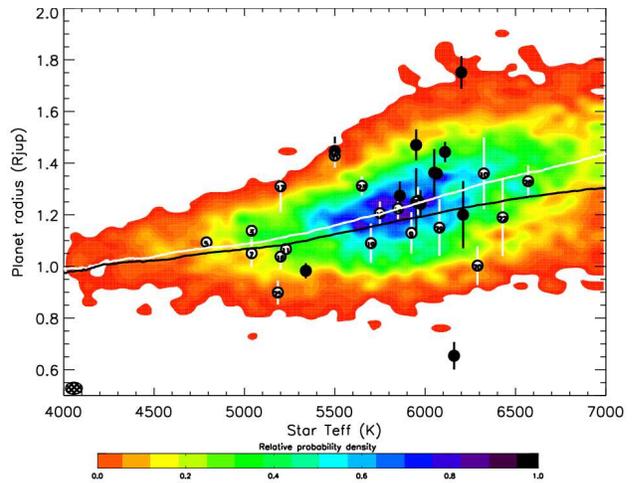}}}
\caption{{\modif Stellar effective temperature versus planetary radius
    of transiting giant planets discovered to date (circles) compared
    to a simulated joint probability density map (contours).} The
  black line is the sliding average of radii in the $[-250 K,+250 K]$
  effective temperature interval for {\it all} simulated transiting
  planets (both below and over the detection threshold). The white
  line is the same average for the {\it detectable} planets in the
  simulation. {\modify The symbols and density map are the same as in fig.~\ref{fig:mass_radius_map}}.}
\label{fig:teff_radius_map}
\end{figure}

  Figure~\ref{fig:teff_radius_map} shows in more detail how stellar
  effective temperature and planetary radius are linked.} We
interpret the {\modif correlation between the two} as the combined
effect of irradiation (visible with the plotted average radius of all
planets with at least one transit event in simulated light curves) and
detection bias (visible with the plotted average radius of simulated
planets detected):
\begin{enumerate}
	\item The planets orbiting bright stars are more
          irradiated. The mean radius of a planet orbiting a warmer
          star is thus higher at a given period. This effect is taken
          into account in our planetary evolution model (see
          \citet{Guillot_2002, Guillot_2006}).
	\item The detection of a planet of a given radius is easier
          for cooler stars since for main sequence stars effective
          temperature and stellar radius are positively correlated.
\end{enumerate}

{\modif We therefore conclude that the effective temperature--planetary
  radius correlation is a {\it consequence} of the physics of the
  problem rather than the cause of the spread in planetary radii. This
  implies that another explanation -- an important variation of the
  planetary composition -- is needed to account for the observed
  radii.

  As in the mass-radius diagram (fig.~\ref{fig:mass_radius_map}),
  there is an outlier at the bottom of
  figure~\ref{fig:teff_radius_map}, HD149026b. As discussed
  previously, this object lies at the boundary of what we could
  simulate, both in terms of masses and amounts of heavy elements, so
  that we do not consider this as significant. It is also presently
  not detectable from a transit survey. Clearly, with more sensitive
  transit surveys, the presence of low-mass planets with large
  fraction of heavy elements compared to hydrogen and helium will
  populate the bottom part of this diagram.

  A last secondary outcome of the study of this diagram concerns the
  possible existence of two groups of planets roughly separated by a
  $T_{\rm eff}=5400$\,K line. We find that the existence of two such
  groups separated by $\sim 200$\,K or more appears serendipitously in
  our model in 10\% of the cases and is therefore
  absolutely not statistically significant. 
} 

\section{Two classes of Hot Jupiters, based on their Safronov numbers?}
\label{sec:safronov}

{\modif According to \cite{Hansen_2007}, the 16 planets discovered at the time
of their study show a bimodal distribution in Safronov numbers, half
of the sample having Safronov numbers $\theta \sim 0.07$ (``class I'') while the
other half is such that $\theta \sim 0.04$ (``class II''). They also
point out that the equilibrium temperatures of the two classes of
planets differ, the class II planets being on average hotter. This is
potentially of great interest because the Safronov number is
indicative of the efficiency with which a planet scatters other bodies
and therefore this division in two classes, if real, may tell us
something about the processes that shaped planetary systems. }

\subsection{{\modif No significant gap between two classes.}}

\begin{figure}
\centerline{\resizebox{9cm}{!}{\includegraphics{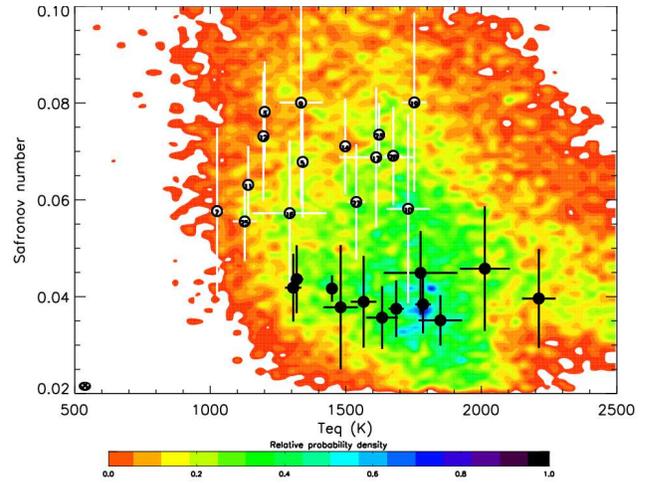}}}
\caption{{\modif Safronov number versus equilibrium temperature of transiting
  giant planets discovered to date (circles) compared to a simulated
  joint probability density map (contours). Open (resp. filled)
  circles correspond to class I (resp. class II) planets. {\modify The
    symbols and density map are the same as in
    fig.~\ref{fig:mass_radius_map}}.} }
\label{fig:safronov_teq}
\end{figure}

{\modif Figure~\ref{fig:safronov_teq} shows how the situation has
  evolved with the new transiting giant planets discovered thus far:
  Although a few planets have narrowed the gap between the two
  ensemble of planets, it is still present and located at a Safronov
  number $\theta\sim 0.05$. The two classes also have mean equilibrium
  temperatures that differ.

  On the other hand, our model naturally predicts a continuous
  distribution of Safronov numbers. A trend is found in which planets
  with high equilibrium temperatures tend to have lower Safronov
  numbers, which is naturally explained by the fact that equilibrium
  temperature and orbital distance are directly linked (remember that
  $\theta=(a/R_{\rm p})\,(M_{\rm p}/M_\star$)).

  We find that our $\theta-T_{\rm eq}$ joint probability density
  function is representative of the observed population, being at
  $0.68\sigma$ from the maximum likelihood (see appendix). A K-S test
  on the Safronov number yields a distance between the observed and
  simulated distributions of 0.163 and a corresponding probability for
  a good match of 0.38, a value that should be improved in future
  models, but that shows that the two ensembles are statistically
  indistinguishable.}

\begin{figure}
\centerline{\resizebox{7cm}{!}{\includegraphics{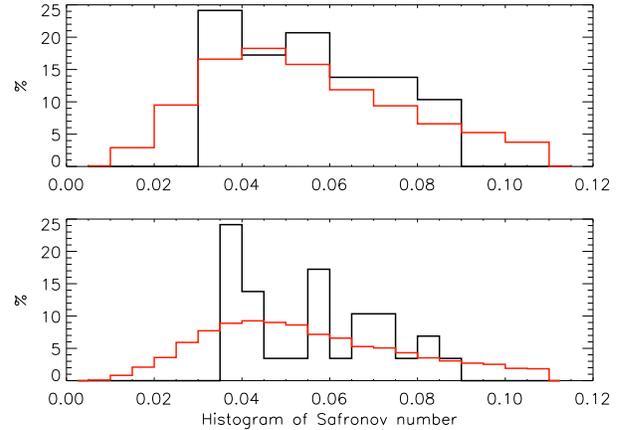}}}
\vspace*{1cm}
\caption{Comparison of the distribution of Safronov number between simulated detections (Red) and real events (Black). Top: histogram with 6 $0.1$-scale columns, Bottom: histogram with 12 $0.05$-scale columns.}
\label{fig:safronov_histogram}
\end{figure}

  Figure~\ref{fig:safronov_histogram}
  compares the histogram of the distribution of Safronov number for
  simulated detections with the histogram of real
  events. Interestingly, although distributions seem different from
  the $0.05$-scale histogram, with a gap appearing in the $0.05-0.055$
  slots, they fit each other while using the $0.1$-scale histogram,
  more appropriate for this low-number statistics analysis (7 intervals for
  31 events).  

\begin{figure}
\centerline{\resizebox{9cm}{!}{\includegraphics{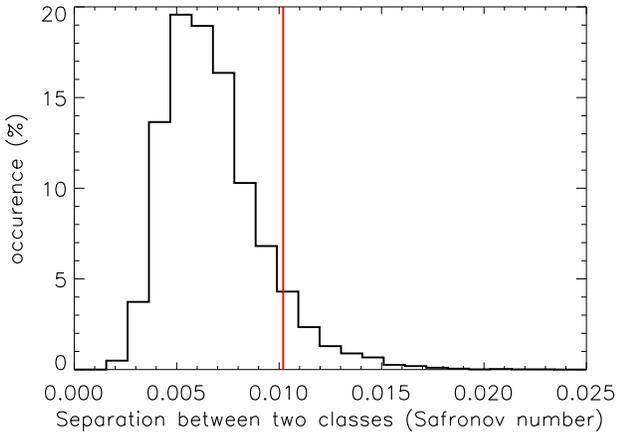}}}
\caption{{\modif Occurrence of the largest observed separation of
    Safronov numbers between two `groups' selected in random draws
    among the model detections sample. The vertical line shows the
    separation ($0.0102$) between the two classes of planets as
    inferred from the observational sample.} 
}
\label{fig:occurrence_of_separation_safronov}
\end{figure}

Figure~\ref{fig:occurrence_of_separation_safronov} shows the
probability to obtain a gap of a given size between the Safronov
numbers of two potential groups of a random draw. {\modif 26 of the
  known transiting Pegasids have their Safronov number between 0 and
  0.1}. Setting a minimum number of 5 planets in each of two classes,
we look for the largest gap between Safronov numbers of a random draw
of 26 simulated Pegasids. For each one of the 10000 Monte-Carlo draws
among the {\modif model detections sample}, we calculate how large is
the most important difference between successive Safronov numbers of
the 26 random draws. We find that a gap of $0.0102$ between two
potential groups is an uncommon event (10 \% of the cases, as 4 \% of
the cases have gaps of this size, and a total of 6 \% of the cases
have larger gaps), yet it is not exceptionally rare. Considering the 7
planet/star characteristics and their many possible combinations, this
level of "rarity" is not statistically significant. 

\begin{figure}
\centerline{\resizebox{9cm}{!}{\includegraphics{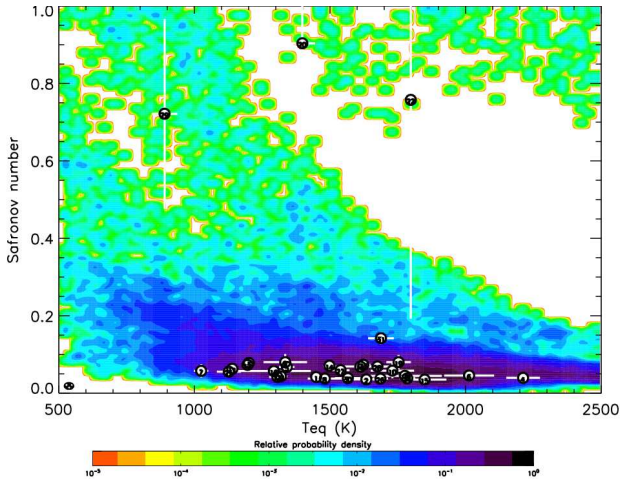}}}
\caption{Same as Figure~\ref{fig:safronov_teq} but for a larger range
  of Safronov numbers. Note that the scale of the color levels is
  logarithmic, in order to emphasize the presence of outliers.}
\label{fig:safronov_teq_extended}
\end{figure}

It is also interesting to consider the few high-Safronov-number
planets discovered {\modif as in
  Figure~\ref{fig:safronov_teq_extended}}. The different gaps in the
diagram are due to our mass vs. period carbon copies of RV planets
that do not uniformly cover the space of parameters. The desert part
in the right edge of the density map is due to the absence of massive
planets in the $[3,15] M_{\rm Jup}$ range at close orbit in the RV
planets. The simulated detections at both high Safronov number and
equilibrium temperature correspond to simulated clones of the planet
HD41004b, with its large mass of 18 $M_{\rm Jup}$ and its very
close-in period of $1.33$ days.

\subsection{{\modif No bimodal distribution visible in other diagrams.}}

\begin{figure}
\centerline{\resizebox{9cm}{!}{\includegraphics{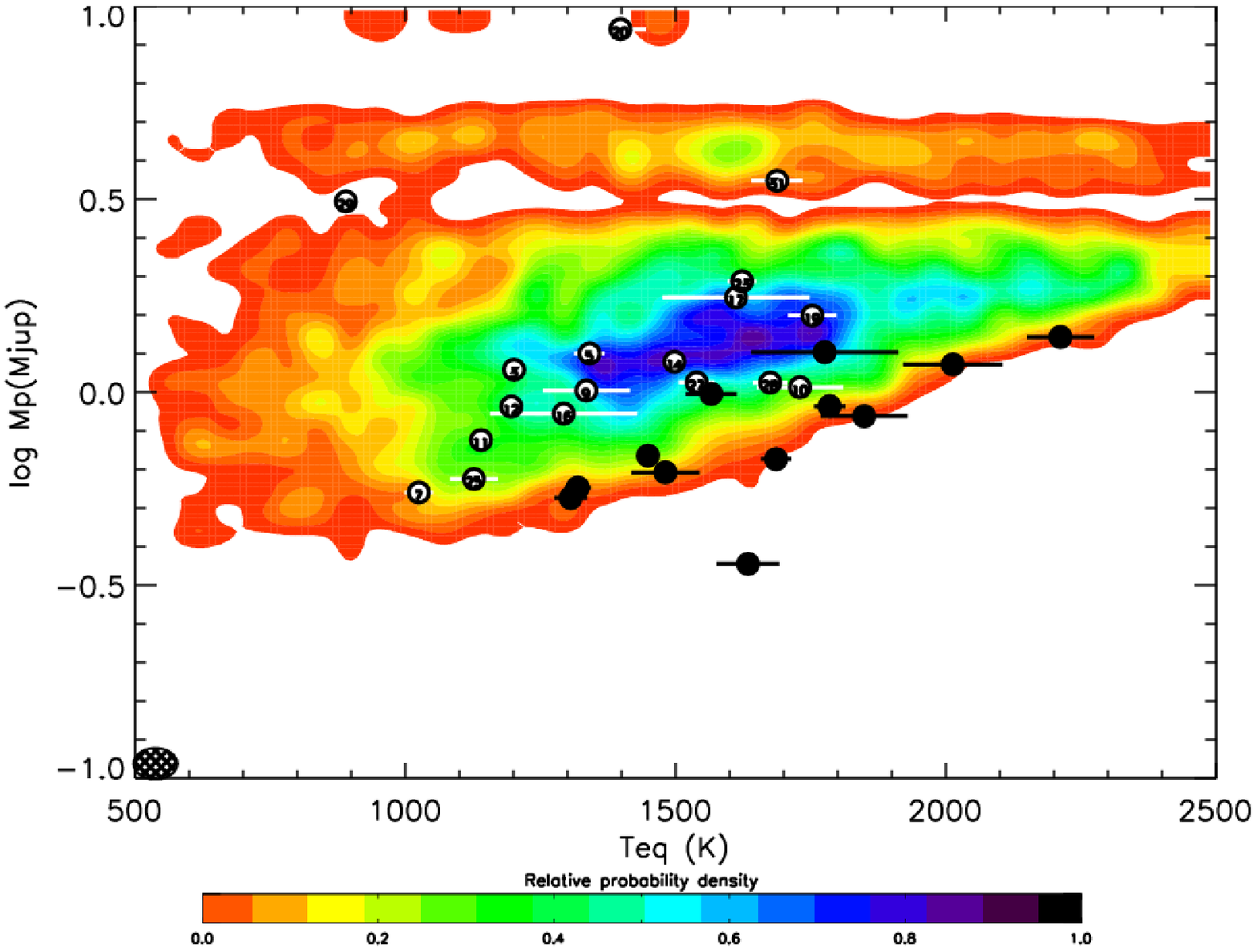}}}
\centerline{\resizebox{9cm}{!}{\includegraphics{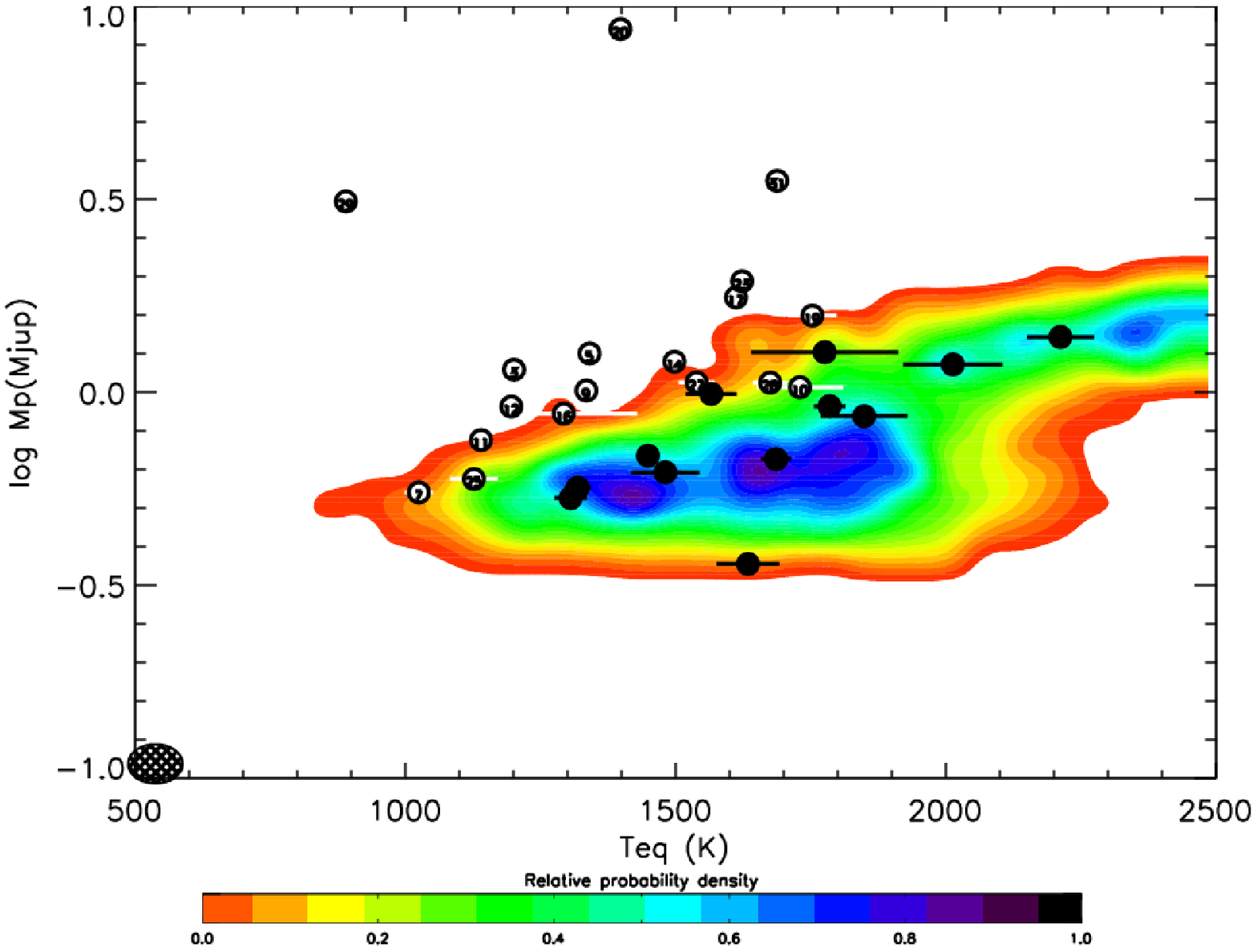}}}
\caption{{\modif Planetary mass versus equilibrium temperature of
    transiting giant planets discovered to date (circles) compared to
    a simulated joint probability density map (contours). {\it Top
      panel:\/} The density map accounts only for simulated planets
    with a Safronov number $\theta > 0.05$ (class~I planets). {\it
      Bottom panel:\/} The density map corresponds only to planets
    with $\theta<0.05$ (class~II planets). {\modify The symbols and density maps are the same as in
    fig.~\ref{fig:mass_radius_map}}.}}
\label{fig:teq_mp_map}
\end{figure}

When plotted as a function of different stellar (effective
temperature, mass, radius) and planetary characteristics (mass,
radius, period, equilibrium temperature), the two potential
{\modif Safronov} classes do not {\modif differ in a significant
  way}. {\modify When plotting our simulated detections as a function of their
Safronov number in different diagrams, the two groups formed by cutting our
model detections sample with a Safronov number cut-off set at $0.05$ 
partly overlap each other on most diagrams}. Here, we choose
to present the planetary mass vs. equilibrium temperature diagram
{\modif which used to provide} a clear separation between the two
populations \citep{Hansen_2007, Torres_2008}. We present in
fig~\ref{fig:teq_mp_map} this diagram as an example of partial overlap
of the class I and class II {\modif detected planets and probability
  density maps}. Contrary to indications based on a smaller sample of
observations, there is no more a clear separation in this diagram
between class I and class II planets.

\subsection{{\modif No correlation between metallicity and Safronov number/class.}}

\citet{Torres_2008} showed that a significant difference could be
observed between the metallicity distributions of the two Safronov
classes. The high-Safronov number class {\modif (class I, $\theta >
  0.05$)} had its host star metallicity centered on $0.0$, and the
{\modif low-Safronov number class (class II)} was centered on
$0.2$. They pointed out that the Safronov numbers for Class I planets
show a decreasing trend with metallicity.

\begin{figure}
\centerline{\resizebox{9cm}{!}{\includegraphics{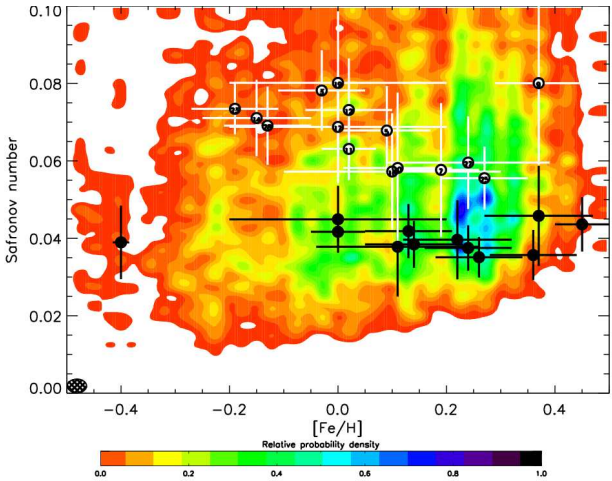}}}
\caption{Safronov number of transiting planets as a function of their host star metallicity. The density map with linear contours comes from the {\modif model detections sample}. Open and filled circles are respectively class I planets {\modify [with Safronov number over $0.05$] and class II planets [with Safronov number below $0.05$]
{\modify Symbols and density plot are the same as in fig.~\ref{fig:mass_radius_map}}.}}
\label{fig:metallicity_safronov_map}
\end{figure}

The two recent discoveries of CoRoT-Exo-1-b ($\rm {\rm [Fe/H]}=-0.4$ and
$\theta=0.038$) and {\modify OGLE TR182-b} ($\rm {\rm [Fe/H]}=0.37$ and $\theta=0.08$)
tend to contradict this argument. Considering the 31 known {\modify giant} planets,
the mean metallicity of stars hosting class I planets is now $\rm
{\rm [Fe/H]}=0.6$, and it is $1.6$ for class II planets.
Figure~\ref{fig:metallicity_safronov_map} shows that although the
metallicity vs. Safronov number distribution of detections we simulate
is a likely result ($0.63 \sigma$ from maximum likelihood), the
potential anticorrelation between $\theta$ and host star {\rm [Fe/H]}
(pointed out by \citet{Torres_2008}) for class I planets is not
present in our simulation, which shows a continouous density map.


\subsection{{\modif No significant gap between two Safronov number classes.}}

Our study has shown us that a separation between two groups of planets linked to
their Safronov number is unlikely for at least two reasons:
\begin{enumerate}
\item The separation between the two groups is marginal. It only
  appears in the Safronov number histogram if the resolution of the
  histogram is high in comparison to the number of events sampled. The
  separation of $\sim 0.01$ between two possible Safronov classes has
  a non-negligible $10\%$ probability to occur serendipitously in our
  distribution which is otherwise continuous. Considering the
  relatively numerous parameters (4 for the star, 3 linked with the
  planet) and their combinations, such a division in two groups
  appears quite likely to occur fortuitously for one such parameter.
\item The separation between the two classes is not present on any
  figures other than the ones involving the Safronov number
  itself. This includes also the separation in metallicity {\modify vs. $\theta$} which is
  not statistically significant, especially given recent discoveries
  of CoRoT-Exo-1b and OGLE-TR-182b . 
\end{enumerate}
On the other hand, we cannot formally rule out the existence of these
two groups of planets. We hence eagerly await other observations of
transiting planets for further tests.

\section{Conclusions}

We have presented a coherent model of a population of stars and
planets that matches within statistical errors the observations of
transiting planets performed thus far. Thanks to new observations, we
have improved on our previous model (Paper~I). In particular, we now
show that with slightly improved assumptions about the metallicity of
stars in the solar neighborhood, the metallicity of stars with
transiting giant planets can be explained without assuming any bias in
period vs. metallicity. 

In order {\modif to validate our model}, we have used a series of
univariate, bivariate and multivariate statistical tests. As the sample of
radial-velocimetry planets and of transiting planets grow, we envision
that with these tools we will be able to much better characterize the
planet population in our Galaxy and its dependence with star population, 
and also test models of planet formation and evolution. 

With today's sample of transiting planets, our model provides a very good match with 
the observations, both when considering {\modif planetary and stellar parameters 
one by one or globally. Our analysis has revealed that the
parameters for the modeled planets are presently statistically
indistinguishable from the observations, although there may be room for improvement
of the model. It should be noted that our underlying assumptions for
the compositions and evolution of planets and for the stellar
populations are relatively simple.} With a larger statistical sample,
tests of these assumptions will be possible and will bring important
constraints on the planet-star distribution in our galactic
neighborhood. The CoRoT mission is expected to be very important in
that respect, especially given the careful determination of the
characteristics of the stellar population that is being monitored. 

Using this method, we have been able to analyze and explain the
different correlations observed between transiting planets
characteristics:
\begin{enumerate}
\item{\it Mass vs. period:\/} One of the first correlations observed
among the planet/star characteristics was the mass vs. period of
close-in RV planets \citep{Mazeh_2005}. Although our model does not
explain it, we confirm with a sample that is now 4 times larger than
at the time of the publication that there is a lack of low-mass planets
($M_p$ $<1\rm M_{\rm Jup}$) on very short periods ($P<2$ days).

\item{\it Surface gravity vs. period:\/} There is an inverse correlation
between the surface gravity and period of transiting planets. We show
that this correlation is caused by the above mass vs. period effect,
and by a lower detection probability for planets with longer periods
and higher surface gravities.

\item{\it Radius vs. stellar effective temperature:\/} Planets around stars with larger
effective temperatures tend to have larger sizes. This is naturally
explained by a combination of slower contraction due to the larger
irradiation and by the increased difficulty in finding planets around
hotter, larger stars. 

\item{Safronov number:\/} \citet{Hansen_2007, Torres_2008} have
  identified a separation between two classes of planets, based on
  their Safrononov number, and visible in different diagrams ($\theta$
  vs. $T_{eq}$ and vs. ${\rm [Fe/H]}$, $M_p$ vs. $T_{eq}$). With recent
  discoveries, this separation is still present in the Safronov number
  distribution, but not anymore in other diagrams. On the other hand,
  our simulation predicts distributions that are continuous, in
  particular in terms of Safronov number.  With this continuous
  distribution, we show that a random draw of 30 simulated planets
  produces two spurious groups separated in Safronov number by a
  distance equal to or larger than the observations in 30\% of the
  cases. The separation is not visible and significant between the two
  classes in any other diagram we plotted. Therefore, we conclude that
  the separation in two classes is not statistically significant but
  is to be checked again with a larger sample of observed
  planets. Interestingly, if on the contrary two classes of Safronov
  numbers were found to exist we would have to revise our 
  model for the composition of planets.
\end{enumerate}

In the next few years, precise analyses of surveys with well-defined
stellar fields and high yields (like CoRoT and Kepler) will allow to
precisely test different formation theories and to link planetary and
stellar characteristics. It should also allow precising the laws
behind the occurrence of planets and their orbital and physical
parameters. Up to now, we have focused on giant planets, but with
larger statistical samples, we hope to be able to extend these kind of
studies to planets of smaller masses which will be intrinsically more
complex because of a larger variety in their compositions (rocks,
ices, gases). {\modify Altogether, this stresses the need for a continuation of radial-velocity and photometric surveys 
for, and follow-up observations of, new transiting planets} 
to greatly
increase the sample of known planets and obtain accurate stellar and
planetary parameters. The goal is of importance: to better understand
what our galactic neighborhood is made of.

\section*{Acknowledgments}

The code used for this work, CoRoTlux, was developed as part of the
CoRoT science program by the authors with major contributions by
Aur\'elien Garnier, Maxime Marmier, Vincent Morello, Martin Vannier,
and help from Suzanne Aigrain, Claire Moutou, St\'ephane Lagarde,
Antoine Llebaria, Didier Queloz, and Fran\c{c}ois Bouchy. {\modif We
  thank F. Pont for many fruitful discussions on the subject, and an
  anonymous referee for a detailed review that helped improve the
  manuscript.} F.F. was funded by a grant from the French {\it Agence
  Nationale pour la Recherche}. This work used Jean Schneider's
exoplanet database {\tt www.exoplanet.eu}, Fr\'ed\'eric Pont's table
of transiting planets characteristics {\tt
  http://www.inscience.ch/transits/} and the Besan\c{c}on model of the
Galaxy at {\tt physique.obs-besancon.fr/modele/} extensively. The
planetary evolution models used for this work can be downloaded at
{\tt www.obs-nice.fr/guillot/pegasids/}.

\bibliography{Fressin_Guillot_Nesta_AA}

\begin{thebibliography}{54}
\expandafter\ifx\csname natexlab\endcsname\relax\def\natexlab#1{#1}\fi

\bibitem[{Abramowitz \& Stegun(1964)}]{Abramowitz_1964}
Abramowitz, M. \& Stegun, I.~A. 1964, Handbook of Mathematical Functions, ninth
  dover printing, tenth gpo printing edn. (New York: Dover)

\bibitem[{{Aldrich} \& {Nelson}(1984)}]{Aldrich_1984}
{Aldrich}, J. \& {Nelson}, F. 1984, {Linear Probability, Logit, and Probit
  Models} (Sage Series on Quantitative Analysis)

\bibitem[{{Alonso} {et~al.}(2008){Alonso}, {Auvergne}, {Baglin}, {Ollivier},
  {Moutou}, {Rouan}, {Deeg}, {Aigrain}, {Almenara}, {Barbieri}, {Barge},
  {Benz}, {Bord{\'e}}, {Bouchy}, {de La Reza}, {Deleuil}, {Dvorak}, {Erikson},
  {Fridlund}, {Gillon}, {Gondoin}, {Guillot}, {Hatzes}, {H{\'e}brard},
  {Kabath}, {Jorda}, {Lammer}, {L{\'e}ger}, {Llebaria}, {Loeillet}, {Magain},
  {Mayor}, {Mazeh}, {P{\"a}tzold}, {Pepe}, {Pont}, {Queloz}, {Rauer},
  {Shporer}, {Schneider}, {Stecklum}, {Udry}, \& {Wuchterl}}]{Alonso_2008}
{Alonso}, R., {Auvergne}, M., {Baglin}, A., {et~al.} 2008, \aap, 482, L21

\bibitem[{{Alonso} {et~al.}(2004){Alonso}, {Brown}, {Torres}, {Latham},
  {Sozzetti}, {Mandushev}, {Belmonte}, {Charbonneau}, {Deeg}, {Dunham},
  {O'Donovan}, \& {Stefanik}}]{Alonso_2004}
{Alonso}, R., {Brown}, T.~M., {Torres}, G., {et~al.} 2004, \apjl, 613, L153

\bibitem[{{Bakos} {et~al.}(2006){Bakos}, {Noyes}, {Latham}, {Cs{\'a}k},
  {G{\'a}lfi}, \& {P{\'a}l}}]{Bakos_2006}
{Bakos}, G., {Noyes}, R.~W., {Latham}, D.~W., {et~al.} 2006, in Tenth
  Anniversary of 51 Peg-b: Status of and prospects for hot Jupiter studies, ed.
  L.~{Arnold}, F.~{Bouchy}, \& C.~{Moutou}, 184--186

\bibitem[{{Bakos} {et~al.}(2007){Bakos}, {Shporer}, {P{\'a}l}, {Torres},
  {Kov{\'a}cs}, {Latham}, {Mazeh}, {Ofir}, {Noyes}, {Sasselov}, {Bouchy},
  {Pont}, {Queloz}, {Udry}, {Esquerdo}, {Sip{\H o}cz}, {Kov{\'a}cs},
  {Stefanik}, {L{\'a}z{\'a}r}, {P{\aap}}, \& {S{\'a}ri}}]{Bakos_2007}
{Bakos}, G.~{\'A}., {Shporer}, A., {P{\'a}l}, A., {et~al.} 2007, \apjl, 671,
  L173

\bibitem[{{Barge} {et~al.}(2008){Barge}, {Baglin}, {Auvergne}, {Rauer},
  {L{\'e}ger}, {Schneider}, {Pont}, {Aigrain}, {Almenara}, {Alonso},
  {Barbieri}, {Bord{\'e}}, {Bouchy}, {Deeg}, {La Reza}, {Deleuil}, {Dvorak},
  {Erikson}, {Fridlund}, {Gillon}, {Gondoin}, {Guillot}, {Hatzes}, {Hebrard},
  {Jorda}, {Kabath}, {Lammer}, {Llebaria}, {Loeillet}, {Magain}, {Mazeh},
  {Moutou}, {Ollivier}, {P{\"a}tzold}, {Queloz}, {Rouan}, {Shporer}, \&
  {Wuchterl}}]{Barge_2008}
{Barge}, P., {Baglin}, A., {Auvergne}, M., {et~al.} 2008, \aap, 482, L17

\bibitem[{{Bouchy} {et~al.}(2005){Bouchy}, {Pont}, {Melo}, {Santos}, {Mayor},
  {Queloz}, \& {Udry}}]{Bouchy_2005}
{Bouchy}, F., {Pont}, F., {Melo}, C., {et~al.} 2005, \aap, 431, 1105

\bibitem[{{Bouchy} {et~al.}(2004){Bouchy}, {Pont}, {Santos}, {Melo}, {Mayor},
  {Queloz}, \& {Udry}}]{Bouchy_2004}
{Bouchy}, F., {Pont}, F., {Santos}, N.~C., {et~al.} 2004, \aap, 421, L13

\bibitem[{{Bouchy} {et~al.}(2008){Bouchy}, {Queloz}, {Deleuil}, {Loeillet},
  {Hatzes}, {Aigrain}, {Alonso}, {Auvergne}, {Baglin}, {Barge}, {Benz},
  {Bord{\'e}}, {Deeg}, {de La Reza}, {Dvorak}, {Erikson}, {Fridlund},
  {Gondoin}, {Guillot}, {H{\'e}brard}, {Jorda}, {Lammer}, {L{\'e}ger},
  {Llebaria}, {Magain}, {Mayor}, {Moutou}, {Ollivier}, {P{\"a}tzold}, {Pepe},
  {Pont}, {Rauer}, {Rouan}, {Schneider}, {Triaud}, {Udry}, \&
  {Wuchterl}}]{Bouchy_2008}
{Bouchy}, F., {Queloz}, D., {Deleuil}, M., {et~al.} 2008, \aap, 482, L25

\bibitem[{{Burke} {et~al.}(2007){Burke}, {McCullough}, {Valenti},
  {Johns-Krull}, {Janes}, {Heasley}, {Summers}, {Stys}, {Bissinger}, {Fleenor},
  {Foote}, {Garc{\'{\i}}a-Melendo}, {Gary}, {Howell}, {Mallia}, {Masi},
  {Taylor}, \& {Vanmunster}}]{Burke_2007}
{Burke}, C.~J., {McCullough}, P.~R., {Valenti}, J.~A., {et~al.} 2007, \apj,
  671, 2115

\bibitem[{{Charbonneau} {et~al.}(2000){Charbonneau}, {Brown}, {Latham}, \&
  {Mayor}}]{Charbonneau_2000}
{Charbonneau}, D., {Brown}, T.~M., {Latham}, D.~W., \& {Mayor}, M. 2000, \apjl,
  529, L45

\bibitem[{{Charbonneau} {et~al.}(2006){Charbonneau}, {Winn}, {Latham}, {Bakos},
  {Falco}, {Holman}, {Noyes}, {Cs{\'a}k}, {Esquerdo}, {Everett}, \&
  {O'Donovan}}]{Charbonneau_2006}
{Charbonneau}, D., {Winn}, J.~N., {Latham}, D.~W., {et~al.} 2006, \apj, 636,
  445

\bibitem[{{Collier Cameron} {et~al.}(2006){Collier Cameron}, {Bouchy},
  {Hebrard}, {Maxted}, {Pollacco}, {Pont}, {Skillen}, {Smalley}, {Street},
  {West}, {Wilson}, {Aigrain}, {Christian}, {Clarkson}, {Enoch}, {Evans},
  {Fitzsimmons}, {Fleenor}, {Gillon}, {Haswell}, {Hebb}, {Hellier}, {Hodgkin},
  {Horne}, {Irwin}, {Kane}, {Keenan}, {Loeillet}, {Lister}, {Mayor}, {Moutou},
  {Norton}, {Osborne}, {Parley}, {Queloz}, {Ryans}, {Triaud}, {Udry}, \&
  {Wheatley}}]{Cameron_2006}
{Collier Cameron}, A., {Bouchy}, F., {Hebrard}, G., {et~al.} 2006, ArXiv
  Astrophysics e-prints

\bibitem[{{Duquennoy} \& {Mayor}(1991)}]{Duquennoy_1991}
{Duquennoy}, A. \& {Mayor}, M. 1991, \aap, 248, 485

\bibitem[{{Fischer} \& {Valenti}(2005)}]{Fischer_2005}
{Fischer}, D.~A. \& {Valenti}, J. 2005, \apj, 622, 1102

\bibitem[{{Fressin} {et~al.}(2007){Fressin}, {Guillot}, {Morello}, \&
  {Pont}}]{Fressin_2007}
{Fressin}, F., {Guillot}, T., {Morello}, V., \& {Pont}, F. 2007, \aap, 475, 729

\bibitem[{{Gillon} {et~al.}(2006){Gillon}, {Pont}, {Moutou}, {Bouchy},
  {Courbin}, {Sohy}, \& {Magain}}]{Gillon_2006}
{Gillon}, M., {Pont}, F., {Moutou}, C., {et~al.} 2006, \aap, 459, 249

\bibitem[{{Gillon} {et~al.}(2007){Gillon}, {Pont}, {Moutou}, {Santos},
  {Bouchy}, {Hartman}, {Mayor}, {Melo}, {Queloz}, {Udry}, \&
  {Magain}}]{Gillon_2007}
{Gillon}, M., {Pont}, F., {Moutou}, C., {et~al.} 2007, \aap, 466, 743

\bibitem[{Greene(2000)}]{Greene_2000}
Greene, W.~H. 2000, {Econometric Analysis}, fourth edition edn. (Prentice Hall
  International, International Edition)

\bibitem[{{Guillot}(2008)}]{Guillot_2008}
{Guillot}, T. 2008, Physica Scripta Volume T, 130, 014023

\bibitem[{{Guillot} {et~al.}(2006){Guillot}, {Santos}, {Pont}, {Iro}, {Melo},
  \& {Ribas}}]{Guillot_2006}
{Guillot}, T., {Santos}, N.~C., {Pont}, F., {et~al.} 2006, \aap, 453, L21

\bibitem[{{Guillot} \& {Showman}(2002)}]{Guillot_2002}
{Guillot}, T. \& {Showman}, A.~P. 2002, \aap, 385, 156

\bibitem[{{Hansen} \& {Barman}(2007)}]{Hansen_2007}
{Hansen}, B.~M.~S. \& {Barman}, T. 2007, \apj, 671, 861

\bibitem[{{Holman} {et~al.}(2006){Holman}, {Winn}, {Latham}, {O'Donovan},
  {Charbonneau}, {Bakos}, {Esquerdo}, {Hergenrother}, {Everett}, \&
  {P{\'a}l}}]{Holman_2006}
{Holman}, M.~J., {Winn}, J.~N., {Latham}, D.~W., {et~al.} 2006, \apj, 652, 1715

\bibitem[{{Knutson} {et~al.}(2007){Knutson}, {Charbonneau}, {Noyes}, {Brown},
  \& {Gilliland}}]{Knutson_2007}
{Knutson}, H.~A., {Charbonneau}, D., {Noyes}, R.~W., {Brown}, T.~M., \&
  {Gilliland}, R.~L. 2007, \apj, 655, 564

\bibitem[{{Konacki} {et~al.}(2003){Konacki}, {Sasselov}, {Torres}, {Jha}, \&
  {Kulkarni}}]{Konacki_2003}
{Konacki}, M., {Sasselov}, D.~D., {Torres}, G., {Jha}, S., \& {Kulkarni}, S.~R.
  2003, in Bulletin of the American Astronomical Society, 1416--+

\bibitem[{{Kov{\'a}cs} {et~al.}(2007){Kov{\'a}cs}, {Bakos}, {Torres},
  {Sozzetti}, {Latham}, {Noyes}, {Butler}, {Marcy}, {Fischer}, {Fern{\'a}ndez},
  {Esquerdo}, {Sasselov}, {Stefanik}, {P{\'a}l}, {L{\'a}z{\'a}r}, {P{\aap}}, \&
  {S{\'a}ri}}]{Kovacs_2007}
{Kov{\'a}cs}, G., {Bakos}, G.~{\'A}., {Torres}, G., {et~al.} 2007, \apjl, 670,
  L41

\bibitem[{{Mandushev} {et~al.}(2007){Mandushev}, {O'Donovan}, {Charbonneau},
  {Torres}, {Latham}, {Bakos}, {Dunham}, {Sozzetti}, {Fern{\'a}ndez},
  {Esquerdo}, {Everett}, {Brown}, {Rabus}, {Belmonte}, \&
  {Hillenbrand}}]{Mandushev_2007}
{Mandushev}, G., {O'Donovan}, F.~T., {Charbonneau}, D., {et~al.} 2007, \apjl,
  667, L195

\bibitem[{{Mazeh} {et~al.}(2005){Mazeh}, {Zucker}, \& {Pont}}]{Mazeh_2005}
{Mazeh}, T., {Zucker}, S., \& {Pont}, F. 2005, \mnras, 356, 955

\bibitem[{{McCullough} {et~al.}(2006){McCullough}, {Stys}, {Valenti},
  {Johns-Krull}, {Janes}, {Heasley}, {Bye}, {Dodd}, {Fleming}, {Pinnick},
  {Bissinger}, {Gary}, {Howell}, \& {Vanmunster}}]{McCullough_2006}
{McCullough}, P.~R., {Stys}, J.~E., {Valenti}, J.~A., {et~al.} 2006, \apj, 648,
  1228

\bibitem[{{Minniti} {et~al.}(2007){Minniti}, {Fern{\'a}ndez}, {D{\'{\i}}az},
  {Udalski}, {Pietrzynski}, {Gieren}, {Rojo}, {Ru{\'{\i}}z}, \&
  {Zoccali}}]{Minniti_2007}
{Minniti}, D., {Fern{\'a}ndez}, J.~M., {D{\'{\i}}az}, R.~F., {et~al.} 2007,
  \apj, 660, 858

\bibitem[{{Nordstr{\"o}m} {et~al.}(2004){Nordstr{\"o}m}, {Mayor}, {Andersen},
  {Holmberg}, {Pont}, {J{\o}rgensen}, {Olsen}, {Udry}, \&
  {Mowlavi}}]{Nordstrom_2004}
{Nordstr{\"o}m}, B., {Mayor}, M., {Andersen}, J., {et~al.} 2004, \aap, 418, 989

\bibitem[{{O'Donovan} {et~al.}(2007){O'Donovan}, {Charbonneau}, {Bakos},
  {Mandushev}, {Dunham}, {Brown}, {Latham}, {Torres}, {Sozzetti}, {Kov{\'a}cs},
  {Everett}, {Baliber}, {Hidas}, {Esquerdo}, {Rabus}, {Deeg}, {Belmonte},
  {Hillenbrand}, \& {Stefanik}}]{O_Donovan_2007}
{O'Donovan}, F.~T., {Charbonneau}, D., {Bakos}, G.~{\'A}., {et~al.} 2007,
  \apjl, 663, L37

\bibitem[{{O'Donovan} {et~al.}(2006){O'Donovan}, {Charbonneau}, {Mandushev},
  {Dunham}, {Latham}, {Torres}, {Sozzetti}, {Brown}, {Trauger}, {Belmonte},
  {Rabus}, {Almenara}, {Alonso}, {Deeg}, {Esquerdo}, {Falco}, {Hillenbrand},
  {Roussanova}, {Stefanik}, \& {Winn}}]{O_Donovan_2006}
{O'Donovan}, F.~T., {Charbonneau}, D., {Mandushev}, G., {et~al.} 2006, \apjl,
  651, L61

\bibitem[{{Pont} {et~al.}(2005){Pont}, {Bouchy}, {Melo}, {Santos}, {Mayor},
  {Queloz}, \& {Udry}}]{Pont_2005}
{Pont}, F., {Bouchy}, F., {Melo}, C., {et~al.} 2005, \aap, 438, 1123

\bibitem[{{Pont} {et~al.}(2004){Pont}, {Bouchy}, {Queloz}, {Santos}, {Melo},
  {Mayor}, \& {Udry}}]{Pont_2004}
{Pont}, F., {Bouchy}, F., {Queloz}, D., {et~al.} 2004, \aap, 426, L15

\bibitem[{{Pont} {et~al.}(2008){Pont}, {Tamuz}, {Udalski}, {Mazeh}, {Bouchy},
  {Melo}, {Naef}, {Santos}, {Moutou}, {Diaz}, {Gieren}, {Gillon}, {Hoyer},
  {Kubiak}, {Mayor}, {Minniti}, {Pietrzynski}, {Queloz}, {Ramirez}, {Ruiz},
  {Shporer}, {Soszy{\'n}ski}, {Szewczyk}, {Szyma{\'n}ski}, {Udry}, {Ulaczyk},
  {Wyrzykowski}, \& {Zoccali}}]{Pont_2008}
{Pont}, F., {Tamuz}, O., {Udalski}, A., {et~al.} 2008, \aap, 487, 749

\bibitem[{{Pont} {et~al.}(2006){Pont}, {Zucker}, \& {Queloz}}]{Pont_2006}
{Pont}, F., {Zucker}, S., \& {Queloz}, D. 2006, \mnras, 373, 231

\bibitem[{{Robin} {et~al.}(2003){Robin}, {Reyl{\'e}}, {Derri{\`e}re}, \&
  {Picaud}}]{Robin_2003}
{Robin}, A.~C., {Reyl{\'e}}, C., {Derri{\`e}re}, S., \& {Picaud}, S. 2003,
  \aap, 409, 523

\bibitem[{{Santos} {et~al.}(2004){Santos}, {Israelian}, \&
  {Mayor}}]{Santos_2004}
{Santos}, N.~C., {Israelian}, G., \& {Mayor}, M. 2004, \aap, 415, 1153

\bibitem[{{Sato} {et~al.}(2005){Sato}, {Fischer}, {Henry}, {Laughlin},
  {Butler}, {Marcy}, {Vogt}, {Bodenheimer}, {Ida}, {Toyota}, {Wolf}, {Valenti},
  {Boyd}, {Johnson}, {Wright}, {Ammons}, {Robinson}, {Strader}, {McCarthy},
  {Tah}, \& {Minniti}}]{Sato_2005}
{Sato}, B., {Fischer}, D.~A., {Henry}, G.~W., {et~al.} 2005, \apj, 633, 465

\bibitem[{{Shporer} {et~al.}(2007){Shporer}, {Tamuz}, {Zucker}, \&
  {Mazeh}}]{Shporer_2007}
{Shporer}, A., {Tamuz}, O., {Zucker}, S., \& {Mazeh}, T. 2007, \mnras, 136

\bibitem[{{Smith} {et~al.}(2006){Smith}, {Collier Cameron}, {Christian},
  {Clarkson}, {Enoch}, {Evans}, {Haswell}, {Hellier}, {Horne}, {Irwin}, {Kane},
  {Lister}, {Norton}, {Parley}, {Pollacco}, {Ryans}, {Skillen}, {Street},
  {Triaud}, {West}, {Wheatley}, \& {Wilson}}]{Smith_2006}
{Smith}, A.~M.~S., {Collier Cameron}, A., {Christian}, D.~J., {et~al.} 2006,
  \mnras, 373, 1151

\bibitem[{{Southworth} {et~al.}(2007){Southworth}, {Wheatley}, \&
  {Sams}}]{Southworth_2007}
{Southworth}, J., {Wheatley}, P.~J., \& {Sams}, G. 2007, \mnras, 379, L11

\bibitem[{{Sozzetti} {et~al.}(2004){Sozzetti}, {Yong}, {Torres}, {Charbonneau},
  {Latham}, {Allende Prieto}, {Brown}, {Carney}, \& {Laird}}]{Sozzetti_2004}
{Sozzetti}, A., {Yong}, D., {Torres}, G., {et~al.} 2004, \apjl, 616, L167

\bibitem[{{Torres} {et~al.}(2007){Torres}, {Bakos}, {Kov{\'a}cs}, {Latham},
  {Fern{\'a}ndez}, {Noyes}, {Esquerdo}, {Sozzetti}, {Fischer}, {Butler},
  {Marcy}, {Stefanik}, {Sasselov}, {L{\'a}z{\'a}r}, {P{\aap}}, \&
  {S{\'a}ri}}]{Torres_2007}
{Torres}, G., {Bakos}, G.~{\'A}., {Kov{\'a}cs}, G., {et~al.} 2007, \apjl, 666,
  L121

\bibitem[{{Torres} {et~al.}(2004){Torres}, {Konacki}, {Sasselov}, \&
  {Jha}}]{Torres_2004}
{Torres}, G., {Konacki}, M., {Sasselov}, D.~D., \& {Jha}, S. 2004, \apj, 609,
  1071

\bibitem[{{Torres} {et~al.}(2008){Torres}, {Winn}, \& {Holman}}]{Torres_2008}
{Torres}, G., {Winn}, J.~N., \& {Holman}, M.~J. 2008, \apj, 677, 1324

\bibitem[{{Udalski}(2003)}]{Udalski_2003}
{Udalski}, A. 2003, Acta Astronomica, 53, 291

\bibitem[{{Winn} {et~al.}(2007{\natexlab{a}}){Winn}, {Holman}, {Bakos},
  {P{\'a}l}, {Johnson}, {Williams}, {Shporer}, {Mazeh}, {Fernandez}, {Latham},
  \& {Gillon}}]{Winn_2007_c}
{Winn}, J.~N., {Holman}, M.~J., {Bakos}, G.~{\'A}., {et~al.}
  2007{\natexlab{a}}, \aj, 134, 1707

\bibitem[{{Winn} {et~al.}(2007{\natexlab{b}}){Winn}, {Holman}, \&
  {Fuentes}}]{Winn_2007}
{Winn}, J.~N., {Holman}, M.~J., \& {Fuentes}, C.~I. 2007{\natexlab{b}}, \aj,
  133, 11

\bibitem[{{Winn} {et~al.}(2007{\natexlab{c}}){Winn}, {Holman}, {Henry},
  {Roussanova}, {Enya}, {Yoshii}, {Shporer}, {Mazeh}, {Johnson}, {Narita}, \&
  {Suto}}]{Winn_2007_b}
{Winn}, J.~N., {Holman}, M.~J., {Henry}, G.~W., {et~al.} 2007{\natexlab{c}},
  \aj, 133, 1828

\bibitem[{{Winn} {et~al.}(2005){Winn}, {Noyes}, {Holman}, {Charbonneau},
  {Ohta}, {Taruya}, {Suto}, {Narita}, {Turner}, {Johnson}, {Marcy}, {Butler},
  \& {Vogt}}]{Winn_2005}
{Winn}, J.~N., {Noyes}, R.~W., {Holman}, M.~J., {et~al.} 2005, \apj, 631, 1215

\end{thebibliography}

\newpage

\cleardoublepage
\section*{Online data}

\section*{Appendix: Statistical evaluation of the model}

\subsection{Univariate tests on individual planet characteristics}

{\modif In this section, we detail the statistical method and tests
  that have been used to validate the model. We first perform basic
tests of our model with simulations repeating multiple times{\modify the number of observations of the}
OGLE survey in order to get $50,000$ detections.  This number was chosen as a compromise between
  statistical significance and computation time}.
Table~\ref{table:mean_sigma} compares the mean values and standard
variations in the observations and in the simulations. The closeness
of the values obtained for the two populations is an indication that
our approach provides a reasonably good fit to the real stellar and
planetary populations, and to the real planet compositions and
evolutions. 

\begin{table*}[btp]
\caption{{\modif Mean values and standard deviations for the system parameters for the observed
transiting planets and our simulated detections}.}
\label{table:mean_sigma}
\centering
\begin{tabular}{ccccccccccc}
\hline\hline\multicolumn{10}{c}{{\bf real planets}} \\
& $M_p$ & $R_p$ & $P$ & $T_{eq}$& $M_{\star}$& $R_{\star}$& $T_{\rm eff}$ &${\rm [Fe/H]}$ & $\theta$ \\
mean & 1.834 &1.235 &3.387 &1510&1.094 &1.164&5764 &0.087 &0.148 \\
$\sigma$ & 2.645& 0.178 &3.540& 300.2& 0.186& 0.304& 464& 0.187& 0.270 \\
\hline\multicolumn{10}{c}{{\bf simulated detections}} \\
& $M_p$ &$R_p$ & $P$ & $T_{eq}$& $M_{\star}$& $R_{\star}$& $T_{\rm eff}$&${\rm [Fe/H]}$ &$\theta$ \\
mean & 1.655& 1.248 &3.217 &1564 &1.073 &1.167&5813 &0.07805&0.129 \\
$\sigma$ & 2.401& 0.186 &2.897& 411.0&
0.195& 0.324& 599& 0.217& 0.270
\\\hline\hline\end{tabular}\end{table*}

However, we do require more advanced statistical tests. First, we
use the so-called Student's t-test to formally compare the mean
values of all characteristics for both types of planets. The
intuition is that, should the model yield simulated planets of
attributes similar to real planets, the average values of these
attributes should not be significantly different from one another.
In other words, the so-called null hypothesis $H_0$ is that the
difference of their mean is zero. Posing $H_0$: $\mu^{r}-\mu^{s}=0$
where superscripts $r$ and $s$ denote real and simulated planets
respectively, and the alternative hypothesis $H_a$ being the
complement $H_a$: $\mu^{r}-\mu^{s}\neq0$, we compute the $t$
statistics using the first and second moments of the distribution of
each planet characteristics as follows:

\begin{equation}
t = \frac{{\left( {\mu_x^{r}  - \mu_x^{s} } \right)}}{{\frac{{s_p
}}{{\sqrt {n_{r}  + n_{s}} }}}},
\end{equation}

\noindent where $x$ is each of the planet characteristics, $n$ is
the size of each sample, and $s_p$ is the square root of the pooled
variance accounting for the sizes of the two population
samples\footnote{The pooled variance is computed as the sum of each
sample variance divided by the overall degree of freedom:
\begin{equation}	
s_p^2  = \frac{{\sum_{i,r} {\left( {x_i  - \mu_x^{r}} \right)^2 +
\sum_{j,s} {\left( {x_j  - \mu_x^{s}}\right)^2 } } }}{{(n_{r} - 1) +
(n_{s}  - 1)}}\end{equation}}. 
The statistics follows a $t$
distribution, from which one can easily derive the two-tailed
critical probability  that the two samples come from one unique
population of planets, i.e. $H_0$ cannot be rejected. 
The results are displayed in Table~\ref{table:agreement2} (Note that
$\theta$ is the Safronov number; other parameters have their usual
meaning). In all cases, the probabilities are larger than
40\%, implying that there is no significant difference in the mean
characteristics of both types of planets. In other words, the two
samples exhibit similar central tendencies.

\begin{table}
\caption{Test of equality of means. Student's $t$ value and critical
probabilities $p$ that individual parameters for both real and simulated
planets have the same sample mean.} 
\label{table:agreement2}\centering
\begin{tabular}{ccc}
\hline \hline
parameter  & {\modif $t$} & {\modif $p$}  \\ 
\hline$M_{\star}$  &     -0.277 &      0.782 \\ 
 ${\rm [Fe/H]}$  &     -0.392 &      0.695 \\ 
{\modify $T_{\rm eff}$}  &      0.707 &      0.480 \\ 
$R_{\star}$  &      0.331 &      0.741 \\ 
      $P$  &     -0.276 &      0.783 \\ 
    $M_p$  &     -0.570 &      0.569 \\ 
    $R_p$  &      0.642 &      0.521 \\ 
 $T_{eq}$  &      0.834 &      0.405 \\ 
 $\theta$  &     -0.585 &      0.559 \\ 
\hline\hline
\end{tabular}
\end{table}

Next, we perform the Kolmogorov-Smirnov test to allow for a more
global assessment of the compatibility of the two populations. This
test has the advantage of being non-parametric, making no assumption
about the distribution of data. This is particularly important since
the number of real planets remains small, which may alter the
normality of the distribution. Moreover, the Kolmogorov-Smirnov
comparison tests the stochastic dominance of the entire
distributions of real planets over simulated planets. To do so, it
computes the largest absolute deviations $D$ between $F_{r}(x)$, the
empirical cumulative distribution function of characteristics $x$
for real planets, and $F_{s}(x)$ the cumulative distribution
function of characteristics $x$ for simulated planets, over the
range of values of $x$: $D = \mathop {\max }\limits_x \left\{
{\left| {F_{real} \left( x \right) - F_{sim} \left( x \right)}
\right|} \right\}$. If the calculated $D$-statistic is greater than
the critical $D^*$-statistic ({\modif provided by the Kolmogorov-Smirnov table --for 31
observations $D^*=0.19$} for a 80\% confidence level and $D^*=0.24$ for a
95\% confidence level--), then one
must reject the null hypothesis that the two distributions are
similar, $H_0: | F_{r}(x)-F_{s}(x) | <D^*$, and accept $H_a:
| F_{r}(x)-F_{s}(x) | \geq D^*$. Table~\ref{table:agreement} shows the
result of the test. The first column provides the D-Statistics, and
the second column gives the probability that the two samples have
the same distribution.

\begin{table}[htbp]
\caption{{\modif Kolmogorov-Smirnov tests. $D$-statistics and critical
probabilities that individual parameters for both real and simulated
planets have the same distribution}.} 
 \label{table:agreement}
 \centering
 \begin{tabular}{ccc}
 \hline
 \hline parameter & {\modif $D$} & {\modif $p$} \\ 
 \hline$M_{\star}$ & 0.154 & 0.492 \\ 
 ${\rm [Fe/H]}$ & 0.161 & 0.438\\ 
 $T_{\rm eff}$& 0.135 & 0.662\\ 
 $R_{\star}$ & 0.141 &  0.612\\ 
 $P$ & 0.145 & 0.572\\ 
 $M_p$ & 0.173 & 0.347\\ 
 $R_p$ & 0.126 & 0.745\\ 
 $T_{eq}$ & 0.180 & 0.303\\ 
 $\theta$ & 0.163 & 0.381\\ 
 \hline\hline
 \end{tabular}
 \end{table}

Again, we  find a good match between the model and observed samples:
the parameters that have the least satisfactory fits are the
planet's equilibrium temperature and the planet mass 
respectively. These values are interpreted as being due to
imperfections in the assumed star and planet populations. It is
important to stress that although the extrasolar planets' main characteristics
(period, mass) are well-defined by the radial-velocity surveys, the
subset of transiting planets is highly biased towards short periods
and corresponds to a relatively small sample in the known
radial-velocity planet population. This explains why the probability
that the planetary mass is drawn from the same distribution in the
model and in the observations is {\modify relatively low}, which may otherwise seem
surprising given that the planet mass distribution would be expected
to be relatively well defined by the radial-velocity measurements.

\subsection{Tests in two-dimensions}\label{sec:2D}
Tests of the adequation of observations and models in two
dimensions,i.e. when considering one parameter as compared to
another one can be performed using the method of maximum likelihood
as described in Paper~I. Table~\ref{table:likelihood} provides
values of the standard deviations from maximum likelihood for
important combinations of parameters. {\modify The second column is a
comparison using all planets discovered by transit surveys, and the third
 column using all known transiting planets (including those discovered by radial
 velocity)}.

\begin{table}
\caption{Standard deviations from maximum likelihood of the
model and  observed transiting planet populations}
\label{table:likelihood}\centering
\begin{tabular}{ccc}
\hline\hline parameter & planets from & all planets \\
& transit surveys &\\
\hline$M_{\star}$ vs. P & 1.19 $\sigma$ & 1.25 $\sigma$\\
$M_{\star}$ vs. P & 1.48 $\sigma$ & 1.62 $\sigma$\\
$R_p$ vs. $M_p$ & 1.70 $\sigma$ & 1.82 $\sigma$\\
P vs. ${\rm [Fe/H]}$ & 1.09 $\sigma$ & 1.09 $\sigma$\\
$R_p$ vs. ${\rm [Fe/H]}$ & 1.61 $\sigma$ & 1.71 $\sigma$\\
g vs. P & 0.61 $\sigma$ & 0.51 $\sigma$\\
$\theta$ vs. $T_{\rm eff}$ & 0.68 $\sigma$ & 0.63 $\sigma$\\
$R_p$ vs. $T_{\rm eff}$ & 1.22 $\sigma$ & 1.45 $\sigma$ \\
  \hline\hline
\end{tabular}
\end{table}

The results are generally good, with deviations not exceeding
$1.82\sigma$. They are also very similar when considering all
planets or only the subset discovered by photometric surveys. This
shows that the radial-velocity and photometric planet characteristics
are quite similar. The mass vs. radius relation shows the highest deviation,
as a few planets are outliers of our planetary evolution model.

\subsection{Multivariate assessment of the performance of the model}

\subsubsection{Principle}

Tests such as the student-t statistics and the Kolmogorov-Smirnov test
are important to determine the adequation of given parameters, but
they do not provide a multivariate assessment of the model. In order
to assess globally the viability of our model we proceed as follows:
We generate a list including {\modif 50,000 ``simulated''} planets and
the 31 {\modif ``observed''} {\modify giant} planets from
Table~\ref{table:transiting_planets}.  
  {\modif This number is necessary to get an accurate multi-variate analysis (see paragraph~\ref{sec:number_required}).}
  A dummy variable $Y$ is generated with value {\modif
  $1$ if the planet is observed, $0$ if the planet is simulated}.


In order to test dependencies between parameters, we have presented in
table~\ref{table:all_correlations} (\S~\ref{sec:stat}) the Pearson
correlation coefficients between each variable including $Y$.  A first
look at the table shows that the method rightfully retrieves the
important physical correlation without any a priori information
concerning the links that exist between the different parameters.  For
example, the stellar effective temperature $T_{\rm eff}$ is positively
correlated to the stellar mass $M_\star$, and radius $R_\star$. It is
also naturally positively correlated to the planet's equilibrium
temperature $T_{\rm eq}$, and to the planet's radius $R_{\rm p}$
simply because evolution models predict planetary radii that are
larger for larger values of the irradiation, all parameters being
equal. Interestingly, it can be seen that although the Safronov number
is by definition correlated to the planetary mass, radius, orbital
period and star mass (see eq.~\ref{eq:safronov}), the largest
correlation parameters {\modify for $\theta$} in absolute value are those related to $M_p$
and $P$ ({\modif as the range of these parameters both vary by more
  than one decade, while $M_{\star}$ and $R_p$ only vary by a factor
  $2$}). Also, we observe that the star metallicity is only correlated
to the planet radius. This is a consequence of our assumption that a
planet's heavy element content is directly proportional to the star's
{\rm [Fe/H]}, and of the fact that planets with more heavy elements are
smaller, all other parameters being equal. The planet's radius is
itself correlated negatively with {\rm [Fe/H]} and positively with $T_{\rm
  eq}$, $M_\star$,$R_\star$ and $T_{\rm eff}$.
Table~\ref{table:all_correlations} also shows the correlations with
the ``reality'' parameter. Of course, a satisfactory model is one in
which there is no correlation between this reality parameter and other
physical parameters of the model. In our case, the corresponding
correlation coefficients are always small and indicate a good match
between the two populations.

Obviously the unconditional probability that a given planet is real
is $\Pr(Y=1)=31/50031\simeq.00062$. Now we wish to know whether this
probability is sensitive to any of the planet characteristics,
controlling for all planet characteristics at once. Hence we model
the probability that a given planet is "real" using the logistic
cumulative density function as follows:
{\modif
\begin{equation}
\label{eq:logistic}
\Pr(Y = 1|{\mathbf{X_i}}) = \frac{{e^{{\mathbf{X_i{b} }}}
}}{{1 + e^{{\mathbf{X_i{b} }}} }}
\end{equation}
}
\noindent where {\modif $\mathbf{X_i}$} is the vector of explanatory
variables (i.e. planet characteristics) for the planet $i$ (real or
simulated), and $\mathbf{{b}}$ is the vector of parameter to be
estimated, and $\mathbf{X_i{b}}\equiv b_0 + \sum_j X_{ij}b_j$, and $b_0$ is a
constant. There are $n$ events to be considered ($i=1..n$) and $m$
explanatory variables ($j=1..m$).  

Importantly, an ordinary least square estimator shall not be used in
this framework, due to the binary nature of the dependent
variables. (Departures from normality and predictions outside the
range $[0;1]$ are the quintessential motivations). Instead, Equation
\ref{eq:logistic} can be estimated using maximum likelihood
methods. The so-called logit specification \citep{Greene_2000} fits
the parameter estimates $\mathbf{{b}}$ so as to maximize the log
likelihood function :
\begin{equation}
\label{likelihood}
\log L({\mathbf{Y}}|{\mathbf{X,{b}}}) = \sum\limits_{i = 1}^n
{y_i\ {{\mathbf{X_i{b} }}}}  - \sum\limits_{i = 1}^n {\log
\left[ {1 + e^{{\mathbf{X_i{b} }}} } \right]}.
\end{equation}
{\modif 
The $\log L$ function is then maximized
choosing $\mathbf{\hat{b} }$ such that {\modif ${\partial \log
    L(y_i,{\mathbf{X_i,\hat{b} }})}/{\partial\mathbf{\hat{b} }}=0$}, using a
Newton-Raphson algorithm. The closer the coefficients
$\hat{b}_1,\hat{b}_2,..,\hat{b}_m$ are to $0$, the closer the model is
to the observations. Conversely, a coefficient that is significantly
different from zero tells us that there is a correlation between this
coefficient and the probability for a planet to be ``real'', i.e. the
model is not a good match to the observations. 

Two features of logistic regression using
maximum likelihood estimators are worth to be mentioned. First, the
value added of the exercise is that the multivariate approach allows
us to hold all other planet characteristics constant, extending the
bivariate correlations to the multivariate case. In other words, we
control for all planet characteristics at once. Second, one can test
whether a given parameter estimate is equal to 0 with the usual null
hypothesis $H_0$: ${b}=0$ versus $H_a$: ${b}\neq 0$. The variance of
the estimator\footnote{The variance of the estimator is provided by
  the Hessian ${\partial^2 \log L(y_i|{\mathbf{X_i,{b}
    }})}/{\partial\mathbf{{b}}\partial\mathbf{{b}^\prime}}$.} is used
to derive the standard error of the parameter estimate. {\modif 
Using equation~\ref{likelihood}}, dividing for each variable
$\hat{{b}}_j$ by the standard error ${\rm s.e.}(\hat{{b}}_j)$ yields
the t-statistics and allows us to test $H_0$. \modif{We note ${\cal P}_j$ 
the probability that a higher value of $t$ would occur by chance . This probability is evaluated for each explanatory variable
  $j$.}  Should our model perform well, we would expect the $t$ value
of each parameter estimate to be null, and the corresponding
probability ${\cal P}_j$ to be close to one. This would imply no
significant association between a single planet characteristics and
the event of being a "real" planet.

Last but not least, the global probability that the model and
observations are compatible can be estimated. To do so, we compute 
the log likelihood obtained when $b_j=0$
for $j=1..m$, where $m$ is the number of variables. Following
eq.~(\ref{likelihood}):
\begin{equation}
\label{likelihood}
\log L({\mathbf{Y}}|1,b_0) = \sum\limits_{i = 1}^n
{y_i b_0}  - \sum\limits_{i = 1}^n {\log
\left[ {1 + e^{b_0}} \right]}.
\end{equation}
The maximum of this quantity is $\log
L_0=n_0\log(n_0/n)+n_1\log(n_{1}/n)$, where $n_0$ is the number of
cases in which $y=0$ and $n_1$ is the number of observations with
$y=1$. $L_0$ is thus the maximum likelihood obtained for a model which
is in perfect agreement with the observations (no explanatory variable
is correlated to the probability of being real).
Now, it can be shown that the likelihood statistic ratio
\begin{equation}
c_{\rm LL}=2 (\log L_1 - \log L_0)
\end{equation}
follows a $\chi^2$ distribution for a number of degrees of freedom $m$
when the null hypothesis is true (\citet{Aldrich_1984}). The
probability that a sum of $m$ normally distributed random variables
with mean $0$ and variance $1$ is larger than a value $c_{\rm LL}$ is:
\begin{equation}
{\cal P}_{\chi^2}= P(m/2,c_{\rm LL}/2),
\end{equation}
where $P(k,z)$ is the regularized Gamma function (e.g. \citet{Abramowitz_1964}). 
${\cal{P}}_{\chi^2}$ is thus the probability that the model
planets and the observed planets are drawn from the same
distribution.

\subsubsection{Determination of the number of model planets required}
\label{sec:number_required}

A problem that arose in the course of the present work was to evaluate
the number of model planets that were needed for the logit evaluation.
It is often estimated that about 10 times more model points than
observations are sufficient for a good tests. We found that
this relatively small number of points indeed leads to a valid
identification of the explanatory variables that are problematic,
i.e. those for which the $\hat{b}$ coefficient is significantly 
different from $0$ (if any). However, the evaluation of the global
$\chi^2$ probability was then found to show considerable statistical
variability, probably given the relatively large number of explanatory
variables used for the study. 

In order to test how the probability ${\cal{P}}_{\chi^2}$ depends on the size
n of the sample to be analyzed, we first generated a very large list
of $N_{0}$ simulated planets with CoRoTlux. We generated by
Monte-Carlo a smaller subset of $n_0\le N_0$ simulated planets that
was augmented of the $n_1=31$ observed planets and computed ${\cal{P}}_{\chi^2}$
using the logit procedure. This exercise was performed 1000 times, and the results are 
shown in fig.~\ref{fig:test_logit}. The resulting ${\cal{P}}_{\chi^2}$ is
found to be very variable for a sample smaller than $\sim 20,000$
planets. As a consequence, we chose to present tests performed for
$n_0=50,000$ model planets. 

\begin{figure}
\label{fig:test_logit}
\centerline{\resizebox{8cm}{!}{\includegraphics{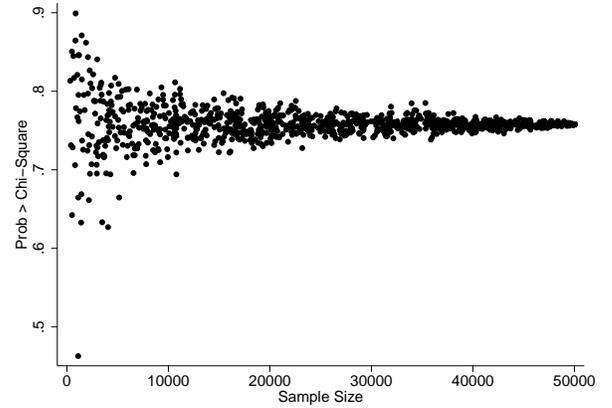}}}
\vspace*{0.5cm}
\caption{Values of the $\chi^2$ probability, ${\cal{P}}_{\chi^2}$ (see
  text) obtained after a logit analysis as a function of the size of
  the sample of model planets $n_0$.} 
\end{figure}




\subsubsection{Analysis of two CoRoTlux samples}

Table~\ref{table:logitres} (see \S~\ref{sec:stat}) reports the
parameter estimates for each of the planet/star characteristics. We
start by assessing the general quality of the logistic regression by
performing the chi-square test. If the vector of planet
characteristics brings no or little information as to which type of
planets a given observation belongs to, we would expect the logistic
regression to perform badly. In technical terms, we would expect the
conditional probability $\Pr(Y = 1|{\mathbf{X}})$ to be equal to the
unconditional probability $\Pr(Y = 1)$. The $\chi^2$ test described
above is used to evaluate the significance of the
model. 

We performed several tests: the first column of results in
table~\ref{table:logitres_10 columns} shows the result of a logit
analysis with the whole series of 9 explanatory variables. Globally,
the model behaves well, with a likelihood statistic ratio $c_{\rm
  LL}=5.8$ and a $\chi^2$ distribution for 9 degrees of freedom
yielding a probability ${\cal{P}}_{\chi^2}=0.758$. When examining
individual variables, we find that the lowest probability derived from
the Student test is that on {\rm [Fe/H]}: ${\cal{P}}_{\rm {\rm [Fe/H]}}=0.164$,
implying that the stellar metallicity is not well reproduced. As
discussed previously, this is due to the fact that several planets of
the observed list have no or very poorly constrained determination of
the stellar {\rm [Fe/H]}, and that a default value of $0$ was then used. 

The other columns in table~\ref{table:logitres_10 columns} show the
result of the logit analysis when removing one variable (i.e. with
only 8 explanatory variables). In agreement with the above analysis,
the highest global probability ${\cal{P}}_{\chi^2}$ is obtained for the
model without the {\rm [Fe/H]} variable. When removing other variables, the
results are very homogeneous, indicating that although the model can
certainly be improved, there is no readily identified problem except
that on {\rm [Fe/H]}. (We hope that future observations will allow for
better constraints on these stars' metallicities).

In order to further test the method, we show in
table~\ref{table:logitres_altered} the results of an analysis in which
the model radii where artificially augmented by 10\%. The
corresponding probabilities are significantly lower: we
find that the model can explain the observations by chance only in
less than 1/10,000. The probabilities for each variable are affected
as well so that it is impossible to identify the culprit for the bad
fit with the 9 variables. However, when removing $R_{\rm p}$ from the
analysis sample, the fit becomes significantly better. (Note that
the results for that column are slightly different than those for the
same column in table~\ref{table:logitres_10 columns} because of the
dependance of $\theta$ with $R_{\rm p}$.)

\begin{table*}[htbp]																							
\caption{Result of the logit analysis for the fiducial model with
  50,000 model planets and 31 observations.}																					
 \label{table:logitres_10 columns}																							
 \centering																							
 \begin{tabular}{cccccccccccc}																							
 \hline
 \hline																							
	&		&	All variables	&	\multicolumn{9}{c}{Missing variable}			\\														
	&		&		&	{\rm [Fe/H]}	&	$T_{eff}$	&	$R_{\star}$	&	$M_{\star}$	&	$P$	&	$R_p$	&	$M_p$	&	$\theta$	&	$T_{eq}$	\\
 \hline																							
{\rm [Fe/H]}	&	${\hat b}_{{\rm [Fe/H]}}$	&	0.415	&		&	0.435	&	0.413	&	0.430	&	0.421	&	0.229	&	0.385	&	0.371	&	0.369	\\
	&	${\cal P}_{{\rm [Fe/H]}}$	&	[0.164]	&		&	[0.148]	&	[0.164]	&	[0.150]	&	[0.157]	&	[0.242]	&	[0.177]	&	[0.196]	&	[0.182]	\\
$T_{eff}$	&	${\hat b}_{T_{\rm eff}}$	&	-0.517	&	-0.579	&		&	-0.541	&	-0.191	&	-0.569	&	-0.563	&	-0.541	&	-0.537	&	-0.618	\\
	&	${\cal P}_{T_{\rm eff}}$	&	[0.417]	&	[0.366]	&		&	[0.372]	&	[0.601]	&	[0.355]	&	[0.378]	&	[0.391]	&	[0.395]	&	[0.298]	\\
$R_{\star}$	&	${\hat b}_{R_{\rm {\star}}}$	&	0.059	&	0.046	&	0.190	&		&	0.212	&	-0.009	&	0.061	&	0.027	&	0.023	&	-0.063	\\
	&	${\cal P}_{R_{\rm {\star}}}$	&	[0.901]	&	[0.924]	&	[0.681]	&		&	[0.609]	&	[0.984]	&	[0.898]	&	[0.953]	&	[0.961]	&	[0.871]	\\
$M_{\star}$	&	${\hat b}_{M_{\rm {\star}}}$	&	0.467	&	0.541	&	-0.001	&	0.511	&		&	0.472	&	0.524	&	0.483	&	0.486	&	0.497	\\
	&	${\cal P}_{M_{\rm {\star}}}$	&	[0.528]	&	[0.468]	&	[0.998]	&	[0.433]	&		&	[0.523]	&	[0.481]	&	[0.512]	&	[0.508]	&	[0.499]	\\
$P$	&	${\hat b}_{P}$	&	-0.236	&	-0.288	&	-0.425	&	-0.199	&	-0.250	&		&	-0.281	&	-0.269	&	-0.356	&	-0.070	\\
	&	${\cal P}_{P}$	&	[0.746]	&	[0.698]	&	[0.605]	&	[0.754]	&	[0.737]	&		&	[0.706]	&	[0.708]	&	[0.618]	&	[0.883]	\\
$R_p$	&	${\hat b}_{R_{\rm p}}$	&	0.305	&	-0.069	&	0.331	&	0.305	&	0.328	&	0.316	&		&	0.261	&	0.246	&	0.241	\\
	&	${\cal P}_{R_{\rm p}}$	&	[0.370]	&	[0.778]	&	[0.332]	&	[0.370]	&	[0.336]	&	[0.352]	&		&	[0.416]	&	[0.456]	&	[0.443]	\\
$M_p$	&	${\hat b}_{M_{\rm p}}$	&	0.329	&	-0.032	&	0.432	&	0.306	&	0.379	&	0.386	&	0.055	&		&	-0.229	&	0.118	\\
	&	${\cal P}_{M_{\rm p}}$	&	[0.726]	&	[0.968]	&	[0.656]	&	[0.737]	&	[0.693]	&	[0.674]	&	[0.947]	&		&	[0.474]	&	[0.876]	\\
$\theta$	&	${\hat b}_{\theta}$	&	-0.904	&	-0.496	&	-1.005	&	-0.879	&	-0.971	&	-1.049	&	-0.625	&	-0.422	&		&	-0.653	\\
	&	${\cal P}_{\theta}$	&	[0.563]	&	[0.706]	&	[0.540]	&	[0.567]	&	[0.548]	&	[0.497]	&	[0.658]	&	[0.410]	&		&	[0.620]	\\
$T_{eq}$	&	${\hat b}_{T_{\rm eq}}$	&	-0.296	&	-0.023	&	-0.520	&	-0.250	&	-0.339	&	-0.169	&	-0.089	&	-0.186	&	-0.150	&		\\
	&	${\cal P}_{T_{\rm eq}}$	&	[0.648]	&	[0.970]	&	[0.414]	&	[0.635]	&	[0.605]	&	[0.744]	&	[0.882]	&	[0.742]	&	[0.801]	&		\\
\multicolumn{12}{c}{overall assessment of the fit}																							\\
\multicolumn{2}{c}{Log likelihood}			&	-257.059	&	-258.123	&	-257.410	&	-257.066	&	-257.275	&	-257.129	&	-257.439	&	-257.126	&	-257.316	&	-257.171	\\
\multicolumn{2}{c}{$c_{\rm LL}$}			&	5.821	&	3.692	&	5.119	&	5.805	&	5.389	&	5.681	&	5.060	&	5.687	&	5.307	&	5.597	\\
\multicolumn{2}{c}{${\cal{P}}_{\chi^2}$}		& 0.758	&	0.884	&	 0.745	&	 0.669	&	 0.715	&	 0.683	&	0.751	&	 0.682	&	 0.724	&	0.692	\\
\hline
\hline
\end{tabular}																							
\end{table*}

\begin{table*}[htbp]																							
\caption{Result of the logit analysis for the altered model ($R_{\rm
    p}$ increased by 10\%) with 50,000 model planets and 31
  observations.}
 \label{table:logitres_altered}																							
 \centering																							
 \begin{tabular}{cccccccccccc}																							
 \hline	
 \hline																						
	&		&	All variables	&	\multicolumn{9}{c}{Missing variable}			\\														
	&		&		&	${\rm [Fe/H]}$	&	$T_{eff}$	&	$R_{\star}$	&	$M_{\star}$	&	$P$	&	$R_p$	&	$M_p$	&	$\theta$	&	$T_{eq}$	\\
 \hline																							
{\rm [Fe/H]}	&	${\hat b}_{{\rm [Fe/H]}}$	&	-0.738	&		&	-0.740	&	-0.737	&	-0.737	&	-0.733	&	0.224	&	-0.607	&	-0.728	&	-0.664	\\
	&	${\cal P}_{{\rm [Fe/H]}}$	&	[0.002]	&		&	[0.002]	&	[0.002]	&	[0.002]	&	[0.002]	&	[0.251]	&	[0.009]	&	[0.002]	&	[0.005]	\\
$T_{\rm eff}$	&	${\hat b}_{T_{\rm eff}}$	&	-0.729	&	-0.713	&		&	-0.742	&	-0.255	&	-0.819	&	-0.573	&	-0.545	&	-0.739	&	-0.308	\\
	&	${\cal P}_{T_{\rm eff}}$	&	[0.260]	&	[0.268]	&		&	[0.231]	&	[0.483]	&	[0.192]	&	[0.366]	&	[0.404]	&	[0.256]	&	[0.618]	\\
$R_{\star}$	&	${\hat b}_{R_{\rm {\star}}}$	&	0.032	&	0.013	&	0.197	&		&	0.247	&	-0.091	&	0.032	&	0.237	&	0.018	&	0.558	\\
	&	${\cal P}_{R_{\rm {\star}}}$	&	[0.945]	&	[0.978]	&	[0.661]	&		&	[0.540]	&	[0.828]	&	[0.945]	&	[0.620]	&	[0.970]	&	[0.149]	\\
$M_{\star}$	&	${\hat b}_{M_{\rm {\star}}}$	&	0.677	&	0.650	&	0.017	&	0.702	&		&	0.684	&	0.532	&	0.557	&	0.667	&	0.598	\\
	&	${\cal P}_{M_{\rm {\star}}}$	&	[0.370]	&	[0.388]	&	[0.966]	&	[0.291]	&		&	[0.363]	&	[0.472]	&	[0.461]	&	[0.377]	&	[0.430]	\\
$P$	&	${\hat b}_{P}$	&	-0.417	&	-0.356	&	-0.664	&	-0.395	&	-0.432	&		&	-0.366	&	-0.393	&	-0.249	&	-1.706	\\
	&	${\cal P}_{P}$	&	[0.585]	&	[0.618]	&	[0.421]	&	[0.565]	&	[0.575]	&		&	[0.622]	&	[0.641]	&	[0.716]	&	[0.037]	\\
$R_p$	&	${\hat b}_{R_{\rm p}}$	&	-1.986	&	-1.264	&	-1.974	&	-1.986	&	-1.985 &	-1.995	&		&	-1.763	&	-1.973	&	-1.796	\\
	&	${\cal P}_{R_{\rm p}}$	&	[0.000]	&	[0.000]	&	[0.000]	&	[0.000]	&	[0.000]	&	[0.000]	&		&	[0.000]	&	[0.000]	&	[0.000]	\\
$M_p$	&	${\hat b}_{M_{\rm p}}$	&	-1.359	&	-0.894	&	-1.350	&	-1.359	&	-1.354	&	-1.305	&	-0.328	&		&	-1.150	&	-1.045	\\
	&	${\cal P}_{M_{\rm p}}$	&	[0.001]	&	[0.019]	&	[0.002]	&	[0.001]	&	[0.001]	&	[0.001]	&	[0.558]	&		&	[0.001]	&	[0.052]	\\
$\theta$	&	${\hat b}_{\theta}$	&	0.384	&	0.271	&	0.461	&	0.376	&	0.387	&	0.193	&	0.021	&	-1.714	&		&	0.338	\\
	&	${\cal P}_{\theta}$	&	[0.359]	&	[0.541]	&	[0.327]	&	[0.347]	&	[0.372]	&	[0.494]	&	[0.976]	&	[0.009]	&		&	[0.633]	\\
$T_{eq}$	&	${\hat b}_{T_{\rm eq}}$	&	1.189	&	0.797	&	0.940	&	1.212	&	1.165	&	1.439	&	-0.009	&	0.603	&	1.202	&		\\
	&	${\cal P}_{T_{\rm eq}}$	&	[0.045]	&	[0.162]	&	[0.100]	&	[0.014]	&	[0.051]	&	[0.001]	&	[0.987]	&	[0.334]	&	[0.054]	&		\\
\multicolumn{12}{c}{overall assessment of the fit}		\\
\multicolumn{2}{c}{Log likelihood}			&	-243.645	&	-247.922	&	-244.341	&	-243.648	&	-244.098	&	-243.872	&	-257.580	&	-246.194	&	-243.872	&	-245.271	\\
\multicolumn{2}{c}{$c_{\rm LL}$}			&	32.647	&	24.094	&	31.256	&	32.643	&	31.743	&	32.194	&	4.778	&	27.551	&	32.194	&	29.395	\\
\multicolumn{2}{c}{${\cal{P}}_{\chi^2}$}			&	 0.000	&	0.002	&	0.000	&	0.000	&	0.000	&	0.000	&	0.781	&	0.001	&	0.000	&	0.000	\\
\hline
\hline
\end{tabular}																							
\end{table*}

\end{document}